\documentclass[a4paper, 11pt]{article}
\usepackage[margin=2cm]{geometry}
\usepackage{amsmath,amssymb,amsthm,amsfonts,array,mathrsfs,ifthen,mathtools}
\usepackage{microtype}
\usepackage{graphicx,epsfig}
\usepackage{lscape}
\usepackage{dcolumn}
\usepackage{eurosym}
\usepackage{enumitem}
\usepackage{newunicodechar}
\newunicodechar{−}{\ensuremath{-}}
\usepackage{multirow}
\usepackage{comment}
\usepackage[normalem]{ulem}

\usepackage[shadow,colorinlistoftodos]{todonotes}

\usepackage[toc,page,header]{appendix}
\usepackage{minitoc}

\usepackage{booktabs, makecell}
\usepackage[flushleft]{threeparttable} 
\usepackage{lscape}
\usepackage{dcolumn}
\usepackage{chngcntr}

\usepackage[semicolon]{natbib}
\DeclareGraphicsExtensions{.pdf,.png,.jpg}
\graphicspath{ {Images/} }
\usepackage{epstopdf}
\usepackage[singlespacing]{setspace}

\usepackage{relsize}

\usepackage{soul} 

\definecolor{myblue}{RGB}{0,30,180}

\usepackage[colorlinks=true,linkcolor=black,
urlcolor=black,
citecolor=myblue,pdftex]{hyperref}

\usepackage[skip=0.333\baselineskip]{caption}
\usepackage{subcaption}
\renewcommand{\arraystretch}{1.2}
\newcolumntype{d}[1]{D{.}{.}{#1}}

\usepackage{float}
\usepackage{tikz}
\usetikzlibrary{arrows.meta, positioning}
\tikzstyle{process} = [
  rectangle, 
  rounded corners=4pt,
  minimum width=3.5cm, 
  minimum height=1.8cm, 
  text centered, 
  draw=black, 
  line width=0.8pt,
  top color=gray!10,
  bottom color=gray!20,
  text width=3.5cm, 
  align=center
]
\tikzstyle{arrow} = [very thick, ->, >=stealth]

\usepackage{xspace}
\newcommand{\Control}{\emph{Urn}\xspace}
\newcommand{\Narrative}{\emph{Story}\xspace}
\newcommand{\NarrativeNI}{\emph{Story No Info}\xspace}
\usepackage[sc]{mathpazo}
\usepackage{sectsty}
\usepackage{titlesec}

\titleformat{\section}
{\normalfont\Large\bfseries\centering\sc} 
{\thesection.}                          
{1em}                                   
{}                                      

\titleformat{\subsection}
{\normalfont\large\bfseries\centering\sc}
{\thesubsection.}
{1em}
{}

\usepackage{authblk}

\begin{document}

\footnotesep=1.2\footnotesep
\baselineskip=1.2\baselineskip

\title{Fooling Yourself: how narratives shape beliefs\thanks{We thank Giorgio Coricelli, I\~{n}igo Iturbe-Ormaetxe, Vladimír Novák, Marcello D'Amato, Giovanni Immordino, and Ivan Soraperra for their helpful comments. 
Paolo Pin acknowledges funding from PRIN grants P20228SXNF and 2022389MRW financed by the Italian Ministry of Research. 
This project has been co-funded by the European Regional Development Fund through the project ``Center for Inequality and Open Society,'' no. CZ.02.01.01/00/23\_025/0008690.\\
Andrea Albertazzi: IMT School for Advanced Studies Lucca, Italy, andrea.albertazzi@imtlucca.it.\\
Paolo Pin: Dipartimento di Economia Politica e Statistica, Universit\`a di Siena, Siena, Italy, and BIDSA,  Universit\`a  Bocconi, Milan, Italy, paolo.pin@unisi.it.\\
Marco Stimolo: Dipartimento di Economia Politica e Statistica, Universit\`a di Siena, Siena, Italy, marco.stimolo@unisi.it.\\
Alessandro Stringhi: Prague University of Economics and Business, Prague, Czechia, alessandro.stringhi@vse.cz.}\\ 
}

\author[a]{Andrea Albertazzi}
\author[b,c]{Paolo Pin}
\author[b]{Marco Stimolo}
\author[d]{Alessandro Stringhi}

\affil[a]{{\small IMT School for Advanced Studies Lucca}} 
\affil[b]{{\small Dipartimento di Economia Politica e Statistica, Universit\`a di Siena}} 
\affil[c]{{\small BIDSA,  Universit\`a  Bocconi}} %
\affil[d]{{\small Prague University of Economics and Business}}

\date{\textit{\today}}
\maketitle
 
\begin{abstract}
\baselineskip = 1.2\baselineskip

\noindent Narratives often combine diagnostic evidence with truthful but nondiagnostic details. In a two-state experiment with known signal likelihoods, we test whether such details distort belief updating. Although Bayes' rule prescribes no revision, nondiagnostic signals induce revisions consistent with considering both states equally likely. Such revisions are most frequent when nondiagnostic signals appear as clues in a story, less frequent in an abstract urn problem, and virtually absent when the same story replaces these clues with explicit no-information messages. Simulations show that these revisions reduce belief accuracy and delay decisions, an effect communicators could exploit without lying.
\medskip

\noindent \textbf{Keywords:} \textit{Narratives; beliefs; nondiagnostic information; uncertainty}.
\\*[0.5\baselineskip]
\noindent \textbf{JEL codes}: C91, D83, D84
\end{abstract}
\clearpage

\noindent

\addtocontents{toc}{\protect\setcounter{tocdepth}{0}}

\clearpage

\section{Introduction}

Many economic decisions --- whether to invest, approve a project, or pursue a course of action --- require extracting relevant evidence from the information available. Yet some details, although factually correct, offer no help in assessing the alternatives. Bayesian updating distinguishes diagnostic evidence, which helps identify the underlying state of the world, from nondiagnostic information, which should leave beliefs unchanged. People may nevertheless revise their beliefs in response to such details, potentially distorting judgments based on informative evidence. Thus, more information need not lead to better decisions.\\

Whether such details sway beliefs may depend on how information is presented. In practice, economic decisions are often framed through narratives rather than in abstract terms. Narratives naturally combine diagnostic evidence with uninformative details by presenting them as connected parts of the same story. This may encourage people to treat each detail as a clue to understanding the situation, even when it provides no evidence about the underlying state \citep{fiske2018social,CHATER2016137,Whitson2008}. Consequently, the same nondiagnostic information may affect beliefs differently depending on whether it is embedded in a narrative or presented in an abstract setting.\\


Building on these observations, we investigate experimentally whether nondiagnostic information, particularly in a narrative context, affects belief updating in a two-state inference problem. We find that, contrary to Bayesian prescriptions, individuals revise their reported beliefs even after receiving information that provides no basis for revision. Specifically, nondiagnostic details can lead people to reset their reported beliefs to exactly 0.5, even when their previous beliefs strongly favored one state over the other. Such resets are more frequent when the details are embedded in a narrative than when statistically equivalent signals are presented in an abstract setting. Within the same narrative, resets are almost absent when nondiagnostic clues are replaced by explicit no-information messages, suggesting that presenting additional details can itself interfere with belief updating. These findings may translate into meaningful economic consequences. Individuals may require more evidence before committing to an investment or undertaking another consequential action.\\

Identifying the effect of nondiagnostic information on beliefs is difficult outside a controlled setting. Apparently irrelevant details may still reveal something about the underlying state, the credibility of a source, or the reliability of earlier evidence. To address this challenge, we conduct a preregistered laboratory experiment in which participants repeatedly report incentivized beliefs about which of two states is true as they observe new signals. These signals are randomly generated according to known state-dependent probabilities and are independent conditional on the true state. Diagnostic signals favor one state over the other, whereas nondiagnostic signals are equally likely in both states and therefore provide no basis for Bayesian revision.\footnote{Nondiagnosticity is a property of the signal-generating process. This does not imply that participants necessarily perceive the signal as irrelevant.} We hold this statistical structure constant across treatments while varying how the inference problem and its signals are presented.\\

We employ a between-subjects design with three treatments. In \Narrative, participants receive diagnostic and nondiagnostic clues within a fictional theft investigation involving two suspects. \NarrativeNI retains the same story and diagnostic clues, while nondiagnostic events occur with the same probability but are presented as explicit no-information messages. Since participants can reconsider their beliefs even when no new clue is provided, this comparison identifies the additional effect of presenting nondiagnostic details. In \Control, participants face a statistically equivalent inference problem involving colored-ball draws, with white balls serving as nondiagnostic signals. Comparing \Narrative and \Control allows us to assess how responses to nondiagnostic signals differ between narrative and abstract settings.\\

Our main finding concerns responses to nondiagnostic signals, which induce substantial departures from Bayesian updating, particularly in \Narrative.\footnote{Responses to diagnostic signals exhibit familiar updating biases, including overreaction and confirmation bias. Following diagnostic signals, the frequency of belief revisions does not differ significantly across treatments, although the magnitude of deviations from the Bayesian benchmark does.} Averaging across participants, the frequency of belief revisions after nondiagnostic signals is 76.3\% in \Narrative, compared with 63.6\% in \Control and 61.7\% in \NarrativeNI. The corresponding mean absolute revisions are 0.142, 0.103, and 0.051, with all pairwise differences statistically significant. More strikingly, these revisions often take the form of \emph{belief resets}: participants report exactly 0.5 after previously reporting a different probability. This pattern is immediately apparent in the raw data, as depicted in Figure~\ref{fig:bubble_nondiagnostic}: in \Narrative, belief reports cluster at exactly 0.5 across a wide range of previous beliefs, including those far from 0.5. A similar, though less pronounced, pattern emerges in \Control. In \NarrativeNI, by contrast, belief reports following a nondiagnostic signal remain much closer to their previous values and exact resets are almost absent.\\

Several complementary analyses reinforce this finding. In \Narrative, \emph{belief resets} remain relatively frequent even when previous beliefs are far from 0.5 or accumulated diagnostic signals strongly favor one state. In \Control, resets become less frequent as beliefs and evidence move away from balance, whereas in \NarrativeNI they are virtually absent. The main qualitative pattern is robust to a broader definition of resets and different specifications. We further examine whether \emph{belief resets} reflect a systematic pattern at the individual level. We classify participants according to alternative updating rules, including two Bayes-like rules that allow for underreaction and overreaction to diagnostic signals but differ in whether nondiagnostic signals leave beliefs unchanged or induce resets. The share of participants classified as Bayes-Reset types is 36.7\% in \Narrative, compared with 21.6\% in \Control and 8.9\% in \NarrativeNI. Conversely, the share classified as Bayes types, whose fitted updating rule prescribes no revision after nondiagnostic signals, is 46.9\%, 65.5\%, and 83.0\%, respectively. Together, these analyses provide converging evidence that \emph{belief resets} constitute a systematic behavioral pattern, particularly pronounced when nondiagnostic information is presented as narrative clues.\\

Belief resets are not merely deviations from the Bayesian benchmark: they can have tangible consequences for economic decisions. By weakening the influence of accumulated diagnostic evidence, resets can reduce belief accuracy and delay decisions that require sufficiently strong beliefs. Such delays may be costly when acquiring additional information is expensive or when postponing action entails foregone opportunities. We illustrate these consequences through simulations that extend the experimentally estimated reset behavior beyond the observed rounds to a sequential decision problem. The simulations show that resets reduce belief accuracy and substantially increase the amount of information needed to form sufficiently strong beliefs.\\

Beyond the consequences illustrated by our simulations, belief resets may also create opportunities for strategic uses of nondiagnostic information in communication. A principal may benefit from preventing others from reaching a firm conclusion, without having to persuade them of the opposite. When the true state is unfavorable to the sender, presenting truthful but nondiagnostic details may induce belief resets that delay recipients from reaching the correct conclusion, without resorting to false claims. Such communication may therefore escape fact-checking focused exclusively on the truthfulness of individual statements, rather than their relevance.\footnote{\cite{mattozzi2026} provide evidence that politicians adjust their communication in response to fact-checking. Fact-checked politicians make fewer factually incorrect statements, but they replace them with unverifiable statements or with no statements at all, rather than with correct ones. Truthful but nondiagnostic details would offer another way to shape beliefs while escaping this kind of scrutiny. Unlike unverifiable statements, they can be checked, and they would pass, because what they lack is relevance rather than accuracy.}\\

If truthful but nondiagnostic details can alter beliefs, ensuring factual accuracy alone may not be sufficient for effective information provision. Courts, for example, regulate the admissibility of evidence based on its relevance and the risk of confusion or prejudice.\footnote{See Federal Rules of Evidence: December 1 (2024), Rules 401--403.} In corporate disclosure, materiality standards distinguish information relevant to investment decisions, while regulatory guidance also emphasizes clear presentation.\footnote{See Rule 12b-2, Title 17 of the Code of Federal Regulations (CFR), \S~240.12b-2; and U.S. Securities and Exchange Commission, \textit{Plain English Disclosure}, Securities Act Release No.~33-7497, January 28, 1998.} Research on investor responses to textual tone further shows that the presentation of information can influence the interpretation of underlying fundamentals \citep{loughran_mcdonald2011, huang_teoh_zhang2014}. Our findings add a distinct concern to these established considerations: nondiagnostic details can induce exact belief resets, rather than merely diluting the influence of diagnostic evidence. In particular, the contrast between \Narrative and \NarrativeNI suggests that filtering such details may help limit belief distortions without removing the narrative context itself.\\

The remainder of this paper is structured as follows. Section~\ref{sec:literature} reviews the relevant literature. Section~\ref{sec:design} describes the experimental design. Section~\ref{sec:results} presents the experimental results and Section~\ref{sec:discussion} provides a discussion. Section~\ref{sec:simulation} investigates the economic implications of our findings. Section~\ref{sec:conclusions} concludes. A series of appendices complements this paper.

\vspace{0.5cm}
\section{Relevant Literature}
\label{sec:literature}

Our paper relates to three main strands of literature. First, it connects to research on nondiagnostic information and the dilution effect, which shows that irrelevant details can weaken judgments based on diagnostic evidence. Second, it contributes to experimental work on departures from Bayesian updating by documenting exact resets of reported beliefs to 0.5 following nondiagnostic signals. Third, it contributes to the economics of narratives by showing that resets are more frequent when nondiagnostic signals are presented as narrative clues than when presented as abstract signals, but are virtually absent when those clues are replaced by explicit no-information messages within the same story.

\paragraph{Dilution effect.}

The literature on the \textit{dilution effect} shows that adding nondiagnostic details alongside diagnostic evidence can reduce the influence of the latter on judgments \citep{nisbett1981dilution,zukier1982dilution}. An important distinction concerns how nondiagnosticity is established. Studies using qualitative descriptions often assess the perceived relevance of individual details, with dilution depending on typicality and conversational context \citep{peters2000typicality,igou2005conversational,kemmelmeier2004separating,kemmelmeier2007conversational,Zukier1984}. In our experiment, nondiagnosticity is defined by the signal-generating process: nondiagnostic clues are equally likely under either state. These clues also avoid directly describing either suspect, limiting their resemblance to the personal attributes commonly studied in this literature.\\

A closer comparison is provided by \citet{TROUTMAN197743} and \citet{labella2004dilution}, who use urn tasks in which nondiagnosticity is defined by the stated urn compositions, but participants observe predetermined, experimenter-selected sequences rather than signals randomly generated according to those compositions. \citet{TROUTMAN197743} find that nondiagnostic samples shift sequential probability judgments toward 0.5. However, \citet{labella2004dilution} do not replicate the dilution effect under sequential belief revision: neutral evidence is largely ignored. Our experiment instead generates signals randomly according to the announced state-dependent probabilities, allowing us to examine belief updating across independently generated signal histories. We document exact belief resets to 0.5 and show that replacing nondiagnostic narrative clues with explicit no-information messages makes these resets rare, even though the narrative context and statistical information structure remain unchanged.

\paragraph{Non-Bayesian updating.}
A vast literature on non-Bayesian updating documents systematic biases in the processing of diagnostic evidence, including underreaction, overreaction, and confirmation bias \citep{BENJAMIN2019,ortoleva2024alternatives}. This literature primarily examines how individuals respond to informative signals, rather than belief revisions following nondiagnostic information. More closely related to our study, \citet{KierenWeber2025} show that uninformative signals induce belief revisions whose direction depends on signal valence. Our findings complement this evidence by documenting a different response to nondiagnostic information. Rather than favoring either state, nondiagnostic narrative clues frequently induce exact belief resets to 0.5, undoing the effect of previously accumulated diagnostic evidence. Moreover, these resets are substantially more frequent when nondiagnostic clues are embedded in a narrative than when they are presented in an abstract setting or replaced by explicit no-information messages. Together, these findings suggest that belief updating in response to nondiagnostic information can depend on how that information is presented, even when its statistical content remains unchanged.\\

Related research documents other ways in which irrelevant, discredited, or previously available information can influence beliefs and judgments. \citet{conlon2026} show that normatively irrelevant cues can shift beliefs by changing which states come to mind, while \citet{Thaler2024} finds that politically congenial yet uninformative signals affect perceived source credibility and posterior beliefs. \citet{musoloff2026} document persistent effects of false stories even after they have been discredited. In financial markets, investors respond to repeated firm news \citep{Tetlock2011}, summary statistics based on previously released data \citep{Gilbert2012}, and recombinations of old news \citep{Fedyk2023}. Related evidence from legal settings shows that vivid but formally irrelevant material can influence judgments \citep{douglas1997impact,bright2006gruesome}. Unlike studies based on financial news or emotionally salient legal evidence, our experiment uses a stylized inference task with known state-dependent signal probabilities, making nondiagnosticity a property of the experimental design.

\paragraph{Narratives and economic beliefs.}
A growing economic literature examines how narratives shape beliefs and decisions. Narratives have been studied as stories that spread and influence aggregate expectations \citep{Shiller2017,Flynn2024,Andre2025}, as causal or interpretive frameworks supporting persuasion and disagreement \citep{Eliaz2020,Schwartzstein2021,aina2025,barron2024}, and as influences on moral and political attitudes, memory, and individual economic decisions \citep{benabou2020narratives,Levy2022,Bursztyn2023,graeber2024stories,morag2025narratives,Han2024,Goetzmann2024}.\\

Our paper highlights a different role of narratives. Rather than persuading individuals toward a particular conclusion, they can hinder the process of learning from evidence. A narrative need not make an alternative explanation more credible to influence decisions; it may suffice to prevent individuals from becoming sufficiently confident in the evidence at hand.

\vspace{0.5cm}
\section{Design and Procedure}\label{sec:design}

We study belief updating in a sequential inference task involving two mutually exclusive states of the world. Participants repeatedly report incentivized beliefs, expressed as probabilities between 0 and 1, as they observe a sequence of signals. The signal-generating process is fully specified: participants know the probability of each signal type under both states, and signals are randomly generated independently conditional on the true state, rather than presented in predetermined sequences. These probabilities distinguish diagnostic signals, which favor one state over the other, from nondiagnostic signals, which are equally likely under both states. The latter have a likelihood ratio of one and therefore provide no basis for Bayesian belief revision. Importantly, diagnosticity is determined by the signal-generating process. However, statistical nondiagnosticity does not necessarily imply behavioral irrelevance.\\

We implement a between-subjects design in which participants are randomly assigned to one of three treatments: \Narrative, \NarrativeNI, or \Control. The statistical structure of the inference problem is identical across treatments, while the context in which the task is framed (narrative or abstract) and the way signals are presented vary.

\begin{itemize}
\item The \Narrative condition is the main treatment of interest. Participants are presented with a fictional story involving a series of thefts in a small town, where one of two suspects is responsible. Diagnostic clues explicitly name one of the suspects, whereas nondiagnostic clues describe details of the investigation without mentioning either suspect. 

\item The \NarrativeNI condition preserves the same narrative environment and diagnostic clues as \Narrative, but participants are told that some rounds may provide no new information. In these rounds, they receive an explicit no-information message instead of a clue and are still asked to report their beliefs.

\item The \Control condition presents the same inferential task in an abstract urn problem. Participants are asked to determine which of two urns is being used to draw a sequence of balls. The urns contain different proportions of two colored balls but the same proportion of white balls. Colored-ball draws are diagnostic, whereas white-ball draws are nondiagnostic.
\end{itemize}

This design allows us to interpret differences in reported belief updating across treatments as effects of how the task and signals are presented. Comparing \Narrative with \NarrativeNI allows us to assess whether presenting nondiagnostic narrative clues affects belief updating, even though these clues provide no more statistical information than explicit no-information messages. This comparison helps distinguish the role of the narrative context itself from the additional effect of presenting nondiagnostic clues within it. Comparing \Narrative with \Control identifies the overall effect of presenting the same statistical inference problem through a narrative with story-based clues rather than an abstract urn task. 
This comparison allows us to assess whether nondiagnostic signals induce greater departures from Bayesian updating when presented as narrative clues than as white ball draws in an abstract urn task.\\

Figure \ref{fig:timeline} summarizes the three parts of the experiment and their order of play.

\vspace{0.5cm}
\subsection{Experimental Design}

\begin{figure}[H]
  \centering \resizebox{0.7\textwidth}{!}{
    \begin{tikzpicture}[node distance=0.5cm and 0.25cm]
    
      \node (stage1) [process] {\textbf{PART 1}\\Introductory\\scenario\\ \ };
      \node (stage2) [process, right=of stage1] {\textbf{PART 2}\\Prior\\elicitation\\$t=0$};
      \node (stage3) [process, right=of stage2] {\textbf{PART 3}\\Posterior\\elicitation\\$t=1,\dots,10$};
    
      \draw [arrow] (stage1) -- (stage2);
      \draw [arrow] (stage2) -- (stage3);

    \end{tikzpicture}}
  \caption{Timeline of the experiment for all conditions.}
  \label{fig:timeline}
\end{figure}

\paragraph{\textbf{PART 1}.} Participants are presented with a fictional scenario designed to establish the context for the belief-updating task. The two introductory scenarios are presented in Table \ref{tab:intro_texts}.

\begin{table}[ht!]
\centering
\caption{Introductory scenarios.}
\begin{tabular}{p{6.5cm}|p{6.5cm}}

\multicolumn{1}{c}{\textbf{\Control}} & \multicolumn{1}{c}{\textbf{\Narrative}/\textbf{\NarrativeNI}}\\
\hline
\multicolumn{2}{c}{\textbf{Main event}} \\
\hline
An experiment was conducted to determine from which of two urns a series of balls was drawn. The experimenters suspect that the chosen urn is either the one containing mostly \emph{Color 1} balls or the one containing mostly \emph{Color 2} balls, but they cannot know for sure. &
A small town has been shaken by a series of thefts. Locals suspect that the culprit could be either Suspect 1, a quiet and reserved man, or \emph{Suspect 2}, known for his outgoing personality, but they have no way to prove it. \\
\hline
\multicolumn{2}{c}{\textbf{Diagnostic Signal}} \\
\hline
Three balls were randomly drawn from the urn in sequence. The first ball was \emph{Color 1}, followed by a second ball of \emph{Color 2}. &
In the past, \emph{Suspect 1}underwent medical treatment for kleptomania. \emph{Suspect 2} had a minor criminal record for petty thefts a few years ago. \\
\hline
\multicolumn{2}{c}{\textbf{Nondiagnostic Signal}} \\
\hline
Finally, the third ball drawn was White. &
A local artist created sketches of both suspects based on eyewitness descriptions, but the portraits looked very much alike, making the identification process even more difficult.\\
\hline
\end{tabular}
\label{tab:intro_texts}
\end{table}

As shown in Table~\ref{tab:intro_texts}, each introductory scenario briefly describes the inference problem and presents two diagnostic signals, one supporting each alternative, together with one nondiagnostic signal. This structure is identical in \Control and \Narrative, with signals presented as ball draws in the former and narrative clues in the latter. The introductory scenario is identical in \Narrative and \NarrativeNI. To reduce the risk that the story and its portrayal of the suspects systematically favor one alternative, we conducted three online validation studies to select a scenario perceived as balanced between the two suspects.\footnote{See Appendix~\ref{validation} for details on the validation procedure.} At the participant level, we also randomized the display position of the true state and the suspect labels in \Narrative and \NarrativeNI, as well as the color labels in \Control.

\paragraph{\textbf{PART 2}.} In this part, participants report their beliefs about which of the two states is true using a continuous slider ranging from 0 (\emph{Color 1}/\textit{Suspect 1}) to 1 (\emph{Color 2}/\emph{Suspect 2}).\footnote{For each slider decision the slider handle was initially hidden, so no default position was displayed until the participant clicked on the slider. This choice was made to prevent any anchoring effect.}
Their responses are incentivized using a quadratic scoring rule that rewards accuracy and penalizes deviations from the true state. Specifically, participants earn up to 100 points if they place the slider exactly on the correct state, with earnings decreasing as their response deviates from it.\footnote{See instructions in Appendix \ref{instructions} for details.} Hence, this part elicits subjects' initial beliefs before the sequential updating task. To ensure comprehension of the task and the visual interface, participants answered a series of control questions before providing their initial probability assessment.

\paragraph{\textbf{PART 3}.} In this part, participants complete ten rounds of belief updating. After observing the signal in each round, they report their beliefs using the slider introduced in Part 2. Each report is incentivized under the same quadratic scoring rule. Before the updating task begins, participants are informed of the signal-generating process, which is identical across treatments. Conditional on the true state, a diagnostic signal favoring that state occurs with probability 0.45, a diagnostic signal favoring the alternative state with probability 0.30, and a nondiagnostic signal with probability 0.25. Participants are explicitly told how these probabilities correspond to the signals presented in their treatment. In \Narrative and \NarrativeNI, diagnostic clues name one suspect. Nondiagnostic signals are presented as clues mentioning neither suspect in \Narrative and as explicit no-information messages in \NarrativeNI. In \Control, colored-ball draws are diagnostic and white-ball draws are nondiagnostic. Participants are informed that nondiagnostic signals have the same probability under both states; their likelihood ratio is therefore one, so Bayesian updating prescribes no belief revision.\\

\subsection{Experimental Procedures}
The experimental design was preregistered at AEA RCT Registry and received ethical approval from the Ethics Committee for Research in the Human and Social Sciences of the University of Siena (Parere n. 17/2024).\footnote{The preregistration is accessible at this link: https://doi.org/10.1257/rct.15672-2.0.} The experiment was conducted at the computer lab of the University of Siena, with a total of 379 participants recruited through the HROOT system \citep{BOCK2014117}. The experimental interface was programmed in oTree \citep{CHEN201688} and distributed via AppLab \citep{PIN2023102666}. Participants were paid for all their probability assessments, therefore their prior and the ten subsequent posterior reports. Earnings were expressed in points and converted into Euros at the end of the experiment at the rate 100 points = 1 Euro. Average payments were 10.84 Euros, including a 4 Euros show-up fee, and sessions lasted on average 30 minutes.\\

Participants were randomly assigned to one of the experimental conditions at the individual level. Additional individual-level randomization concerned the order in which the true and alternative states were displayed, the names assigned to suspects in \Narrative and \NarrativeNI, and the colors used for the two alternatives in \Control. In each round, participants were reminded of the history of signals received up to that round.

\vspace{0.5cm}
\section{Results}\label{sec:results}

In this section, we present the main findings of the experiment. We first show that participants in the \Narrative treatment deviate more from the Bayesian benchmark than participants in \NarrativeNI and \Control. A key source of this difference is their response to nondiagnostic signals: in \Narrative, subjects revise their beliefs more often precisely when the Bayesian benchmark prescribes no update. We next show that nondiagnostic signals can induce exact resets to 0.5 even when prior beliefs or accumulated evidence strongly favour one state. Finally, we classify participants according to the updating rule that best fits their reported beliefs and find that \Narrative increases the share of participants classified as following a rule that resets beliefs to 0.5 after nondiagnostic signals.

\subsection{Departures from Bayesian Updating}

To benchmark belief updating, we compute Bayesian posteriors using participants' stated beliefs from the previous round as priors. This allows us to measure deviations from Bayesian updating in response to each new signal, without attributing deviations accumulated in earlier rounds to the current update.\\

\begin{table}[htbp]
\centering \footnotesize
\caption{Absolute deviations from Bayesian benchmark.}
\begin{tabular}{lccccccc}
\toprule
\multirow{2}{*}{Treatment}
& \multicolumn{3}{c}{Deviations}
& & \multicolumn{2}{c}{Update rate (\%)}
&  \\
\cmidrule(lr){2-4} \cmidrule(lr){5-7}
& All & Diagnostic & Nondiagnostic & & Diagnostic& Nondiagnostic& N\\
\midrule
\Control & 0.121 (0.098) & 0.127 (0.092) & 0.103 (0.148) & & 84.3 (24.2) & 63.6 (39.6) & 139 \\
\Narrative & 0.155 (0.099) & 0.160 (0.107) & 0.142 (0.139) & & 84.2 (21.8) & 76.3 (33.0) & 128 \\
\NarrativeNI & 0.095 (0.064) & 0.108 (0.070) & 0.051 (0.080) & & 86.5 (18.1) & 61.7 (41.5) & 112 \\
\bottomrule
\end{tabular}
\caption*{\footnotesize\textit{Note:} The table reports means of individual-level absolute deviations from the Bayesian theoretical benchmark, for each treatment. The first column includes all signals; the second focuses on diagnostic signals; and the third focuses on nondiagnostic signals only. The fourth and fifth columns report the average subject-level frequency of belief revisions following diagnostic and nondiagnostic signals, respectively. Standard deviations are reported in parentheses. The final column reports the number of subjects in each treatment. The prior round is excluded.}
\label{tab:abs_deviation_summary}
\end{table}

Table~\ref{tab:abs_deviation_summary} reports absolute deviations from the Bayesian benchmark and the frequency of belief revisions by treatment and signal diagnosticity. Mean absolute deviations are largest in \Narrative (0.155), followed by \Control (0.121) and \NarrativeNI (0.095). To account for repeated observations within participants, all nonparametric tests use participant-level averages. A Kruskal--Wallis (KW) test rejects equality of the distributions of overall deviations across treatments $(p<0.001)$. Pairwise Mann--Whitney--Wilcoxon (MWW) tests indicate larger deviations in \Narrative than in both \Control and \NarrativeNI ($p<0.001$ for both comparisons).\\

Disaggregating the analysis by signal diagnosticity clarifies the source of this pattern. Following diagnostic signals, the frequency of belief revisions does not differ significantly across treatments (KW, $p=0.781$). However, the magnitude of deviations from the Bayesian benchmark is higher in \Narrative (MWW, $p=0.006$ and $p<0.001$ against \Control and \NarrativeNI, respectively), whereas \Control and \NarrativeNI do not differ significantly (MWW, $p=0.164$).\\

After nondiagnostic signals, by contrast, any belief revision constitutes a departure from Bayesian updating. Deviations are 0.142 in \Narrative, compared with 0.103 in \Control and 0.051 in \NarrativeNI, and all pairwise comparisons are statistically significant (MWW, $p<0.01$). Revisions are also more frequent in \Narrative: subjects update in 76.3\% of cases, compared with 63.6\% in \Control and 61.7\% in \NarrativeNI. Equality of update frequencies is rejected (KW, $p=0.014$), with \Narrative significantly higher than both \Control (MWW, $p=0.011$) and \NarrativeNI (MWW, $p=0.011$), while \Control and \NarrativeNI are statistically indistinguishable (MWW, $p=0.826$).\\

This evidence shows that departures from the Bayesian benchmark are largest in \Narrative, particularly after nondiagnostic signals.\footnote{Responses to diagnostic signals also exhibit systematic departures from Bayesian updating across all treatments, including overreaction and confirmation bias. Appendix~\ref{app:pre-reg} reports this analysis.} We next examine how nondiagnostic signals change reported beliefs.

\subsection{Belief revisions after nondiagnostic signals}

\begin{figure}[H]
    \centering
\includegraphics[width=0.95\linewidth]{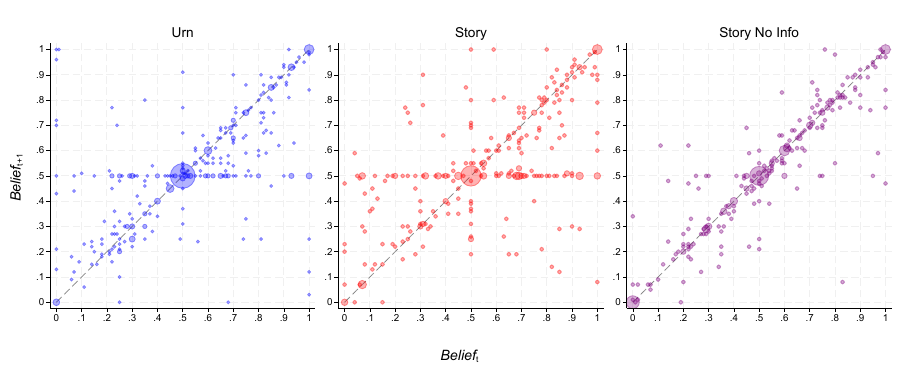}
    \caption{Updating after nondiagnostic signals.\\{\footnotesize \textit{Note}: The bubble plots illustrate belief updating following a nondiagnostic signal. Each tile shows how beliefs evolve from round $t$ (x-axis) to round $t+1$ (y-axis) for each treatment. The size of the marker indicates the number of data points for that specific coordinate, where the smallest marker refers to one observation only.}} \label{fig:bubble_nondiagnostic}
\end{figure}

Figure~\ref{fig:bubble_nondiagnostic} plots current reported beliefs against previous-round beliefs for rounds in which participants observe a nondiagnostic signal. Observations on the 45-degree line indicate unchanged beliefs, as prescribed by Bayes' rule; observations above or below it indicate upward or downward revisions, respectively.\footnote{Figure~\ref{fig:bubbleplots} in Appendix~\ref{app:additional_results} also reports responses to diagnostic signals.}

The figure reveals a distinctive pattern in \Narrative: after nondiagnostic clues, reported beliefs cluster along the horizontal line at $0.5$, often following previous reports far from this value. We refer to revisions from a different probability to exactly $0.5$ as \emph{belief resets}. A similar pattern is visible in \Control, though resets are less frequent. In \NarrativeNI, exact resets are rare, and observations are more clustered along the 45-degree line.\\

\begin{figure}[H]
    \centering
\includegraphics[width=1\linewidth]{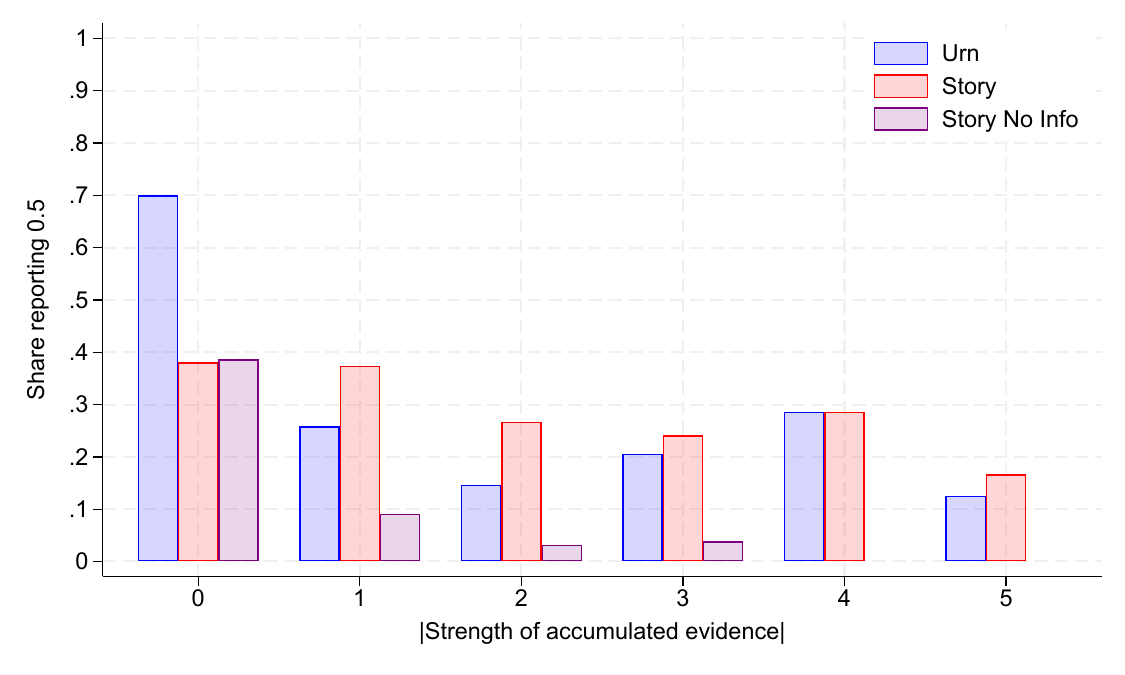}

\caption{Reporting exactly $0.5$ after nondiagnostic signals.\\
{\footnotesize\textit{Note:} The figure shows, for each experimental condition, the share of reports equal to exactly $0.5$ following a nondiagnostic signal, grouped by the absolute imbalance in accumulated diagnostic evidence. All nondiagnostic rounds are included, including unchanged reports of $0.5$.}}
\label{fig:shares_midpoint_strength}
\end{figure}

Figure~\ref{fig:shares_midpoint_strength} further examines this pattern by showing the share of reports equal to exactly $0.5$ after nondiagnostic signals, grouped by the absolute imbalance in accumulated diagnostic evidence. This imbalance is determined by the randomly generated signal history. Zero indicates balanced diagnostic evidence, while positive values count the excess of diagnostic signals favoring one state over those favoring the alternative one.\footnote{Figure~\ref{fig:shares_midpoint} in Appendix~\ref{app:additional_results} instead groups observations by the absolute distance of the previous-round belief from $0.5$. Since these beliefs are themselves affected by treatment, comparisons within each group are descriptive rather than causal.}

When accumulated diagnostic evidence favors one state, reports of exactly $0.5$ are more frequent in \Narrative than in \NarrativeNI (MWW, $p<0.001$) and \Control, although the latter difference is not statistically significant at the 5\% level (MWW, $p=0.069$). Such reports are also more frequent in \Control than in \NarrativeNI (MWW, $p<0.001$). When diagnostic evidence is balanced, by contrast, reports of $0.5$ are most frequent in \Control.\\

We next compare exact reset incidence across treatments and levels of accumulated diagnostic evidence. Using all nondiagnostic rounds, we define an indicator equal to one when a belief changes from a value different from $0.5$ to exactly $0.5$, and zero otherwise, including unchanged reports of $0.5$. We estimate a linear probability model with treatment indicators, evidence-category indicators, their interactions, and round fixed effects, clustering standard errors at the subject level. Evidence categories distinguish whether one state has received zero, one, two, three, four, or five or more diagnostic signals in excess of the other. Table~\ref{tab:responses_by_evidence} reports pairwise treatment differences in percentage points within each category and a joint test of whether these differences vary across categories.\\

\begin{table}[H]
\centering
\caption{Exact belief resets by accumulated evidence}
\label{tab:responses_by_evidence}
\footnotesize
\setlength{\tabcolsep}{8pt}
\renewcommand{\arraystretch}{1.15}
\begin{tabular}{lccc}
\toprule
Evidence imbalance
    & Story $-$ Urn
    & Story $-$ Story No Info
    & Urn $-$ Story No Info \\
\midrule
0 (balanced)
    & $4.17\;(0.485)$
    & $15.85^{***}\;(0.004)$
    & $11.67^{**}\;(0.022)$ \\
1
    & $6.30\;(0.270)$
    & $15.39^{***}\;(0.005)$
    & $9.09^{*}\;(0.065)$ \\
2
    & $8.51\;(0.172)$
    & $17.64^{***}\;(0.001)$
    & $9.13^{**}\;(0.048)$ \\
3
    & $2.81\;(0.796)$
    & $23.05^{***}\;(0.004)$
    & $20.24^{***}\;(0.006)$ \\
4
    & $9.51\;(0.506)$
    & $28.01^{**}\;(0.022)$
    & $18.50^{**}\;(0.020)$ \\
5+
    & $5.97\;(0.709)$
    & $15.44\;(0.227)$
    & $9.47\;(0.335)$ \\
\midrule
Joint interaction test ($p$)
    & \multicolumn{3}{c}{$0.933$} \\
Round fixed effects
    & \multicolumn{3}{c}{Yes} \\
Observations
    & \multicolumn{3}{c}{952} \\
Participants
    & \multicolumn{3}{c}{357} \\
\bottomrule
\end{tabular}

\medskip
\begin{minipage}{0.98\linewidth}
\footnotesize

\textit{Notes:} Entries report treatment differences in exact reset probabilities (percentage points), with two-sided $p$-values in parentheses. Estimates come from an LPM with treatment, categorical evidence imbalance, their interactions, and round fixed effects; standard errors are clustered by participant. Evidence imbalance is the absolute difference in diagnostic signal counts favoring each state. Exact resets are revisions from a belief other than $0.5$ to exactly $0.5$. All nondiagnostic rounds are included. The joint test covers treatment--evidence interactions. The 5+ category pools 28 observations. $^{*}p<0.10$, $^{**}p<0.05$, $^{***}p<0.01$; $p$-values are unadjusted for multiple testing.
\end{minipage}
\end{table}

Table~\ref{tab:responses_by_evidence} shows that resets are more frequent in \Narrative than in \NarrativeNI, whether accumulated evidence is balanced or favors one state. The difference is 15--28 percentage points and statistically significant for evidence imbalances of one through four. Resets are also more frequent in \Control than in \NarrativeNI across these categories, although the evidence is weaker for an imbalance of one ($p=0.065$). Differences between \Narrative and \Control are positive but not statistically significant. None of the comparisons is significant in the highest imbalance category, which contains only 28 observations. The joint test provides no evidence that treatment differences change with evidence imbalance ($p=0.933$). Thus, the clearest finding is that replacing nondiagnostic clues with explicit no-information messages within the same story reduces reset incidence, even when accumulated evidence favors one state.\\


We next examine exact belief resets to $0.5$ following nondiagnostic signals and how their probability varies with previous reported beliefs and across treatments. We exclude observations in which the previous belief is already $0.5$. Because previous beliefs may themselves be affected by treatment, we interpret these conditional comparisons as descriptive rather than causal. The dependent variable indicates whether the updated report equals exactly $0.5$ and is defined as follows:

\[
Y_{it}=\mathbf{1}\{b_{it}=0.5\}.
\]

\noindent We estimate the following model:

\begin{equation}\label{eq:toward}
\begin{aligned}
Y_{it}
=
&\ \alpha
+\beta_1 \Control_i
+\beta_2 \NarrativeNI_i
+\gamma d_{i,t-1} +\delta_1 \Control_i \times d_{i,t-1} \\ &
+\delta_2 \NarrativeNI_i \times d_{i,t-1}
+\eta_{a_{it}}
+\lambda_t
+u_i
+\varepsilon_{it}.
\end{aligned}
\end{equation}

\vspace{0.5cm}

Here, $d_{i,t-1}=|b_{i,t-1}-0.5|$ measures the distance of the previous reported belief from $0.5$, and \Narrative is the reference treatment. We include fixed effects for the absolute imbalance in accumulated diagnostic evidence, $\eta_{a_{it}}$, round fixed effects, $\lambda_t$, and a subject-level random intercept, $u_i$. We exclude observations with $b_{i,t-1}=0.5$, so that a subsequent report of $0.5$ represents a change in belief rather than an unchanged report. The coefficient $\gamma$ captures the association between prior distance from $0.5$ and the latent propensity to reset in \Narrative; $\delta_1$ and $\delta_2$ indicate how the relationship between prior distance from $0.5$ and the latent propensity to reset differs in \Control and \NarrativeNI, relative to \Narrative. We interpret the results using predicted reset probabilities across treatments and prior distances.\\

\begin{figure}[H]
    \centering
\includegraphics[width=1\linewidth]{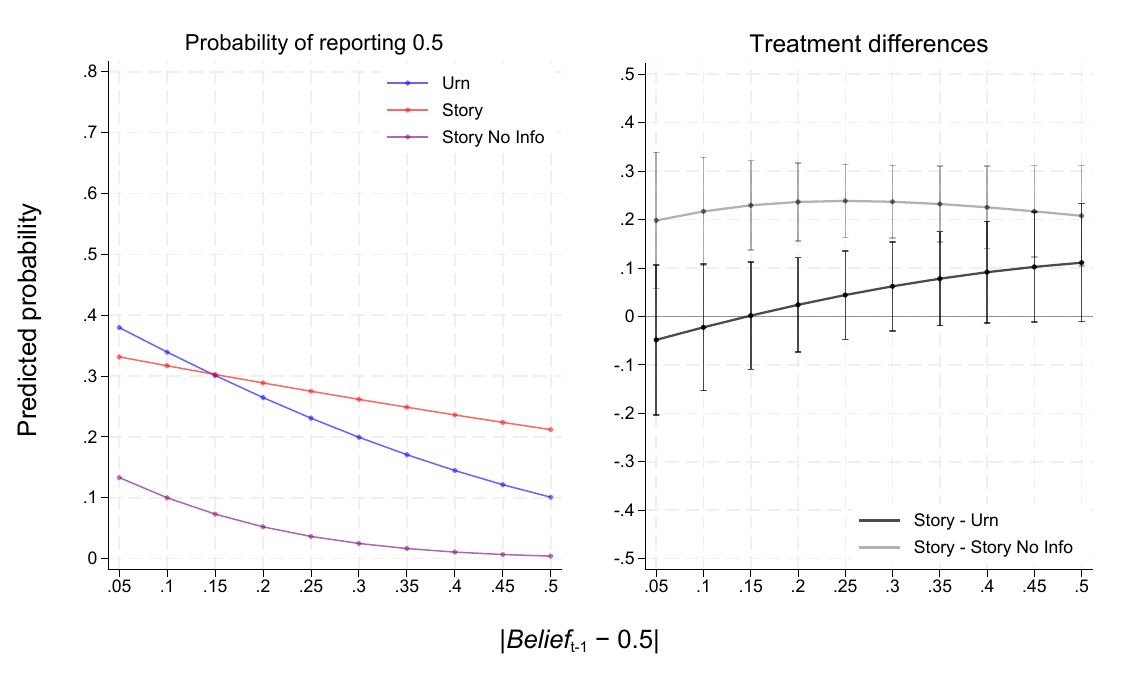}
    \caption{Predicted probability of exact belief resets.\\{\footnotesize\textit{Note:} The left panel shows the predicted probability of reporting a belief of exactly 0.5 after observing a nondiagnostic signal for all treatment conditions. The right panel shows the predicted treatment differences for the relevant treatment comparisons. Vertical bars represent 95\% confidence intervals.}}\label{fig:prob_midpoint}
\end{figure}

Figure~\ref{fig:prob_midpoint} plots predicted reset probabilities by treatment against the distance of the previous reported belief from $0.5$ (left panel), alongside between-treatment differences and their 95\% confidence intervals (right panel).

The figure shows that as previous beliefs move farther from $0.5$, reset probabilities decrease more rapidly in \Control than in \Narrative. Thus, resets remain relatively more frequent in \Narrative even when previous beliefs strongly favor one state. In \NarrativeNI, reset probabilities remain low regardless of the previous belief.\\

The right panel shows that, when previous beliefs are close to $0.5$, predicted reset probabilities are slightly lower in \Narrative than in \Control. This ordering reverses as previous beliefs move farther from $0.5$, although the differences are generally not statistically significant at the 5\% level. By contrast, predicted reset probabilities are consistently higher in \Narrative than in \NarrativeNI. These results show that nondiagnostic signals can prompt exact belief resets to $0.5$, which remain relatively frequent in \Narrative even when previous beliefs strongly favor one state.\footnote{Appendix~\ref{app:additional_results} reports the regression estimates (Table~\ref{reg:prob}) and robustness checks using a broader outcome definition (Figure~\ref{fig:prob_midpoint_robust}) and a quadratic in $d_{i,t-1}$ fully interacted with treatment (Figure~\ref{fig:prob_midpoint_quad}). Both checks preserve the broad pattern: predicted probabilities of resets are higher in \Narrative than in \NarrativeNI throughout, and higher than in \Control over much of the range of prior distances from $0.5$.}

\subsection{Persistence of Belief Resets}

The evidence above shows that nondiagnostic signals can induce reported belief resets to $0.5$. These resets may carry into subsequent updating or reflect temporary reporting responses, after which participants resume updating from their pre-reset beliefs. To distinguish these possibilities, we examine resets immediately followed by a diagnostic signal. We compare the subsequent belief report with two Bayesian benchmarks: a persistent benchmark that uses $0.5$ as the prior and a transitory benchmark that uses the belief reported before the reset.\\

We restrict attention to nondiagnostic rounds $t$ in which $b_{i,t}=0.5$ and $b_{i,t-1}\neq0.5$, immediately followed by a diagnostic signal at $t+1$. For each event, we construct two predictions for the subsequent report, $b_{i,t+1}$, by applying Bayes' rule to the same diagnostic signal using different priors. The persistent benchmark, $b^{Persistent}_{i,t+1}$, uses the reset belief of $0.5$ as the prior, whereas the transitory benchmark, $b^{Transitory}_{i,t+1}$, uses the pre-reset belief, $b_{i,t-1}$.\\

Each benchmark yields a distinct, falsifiable prediction, conditional on Bayesian updating after the subsequent diagnostic signal. We assess departures from both benchmarks separately. Rejecting the transitory benchmark would provide evidence against participants simply reinstating their pre-reset beliefs and updating them according to Bayes' rule. Failure to reject the persistent benchmark would suggest compatibility with updating from $0.5$.\\

For each reset followed by a diagnostic signal, we compute
signed deviations from the two benchmarks:
\begin{equation}
\begin{aligned}
Y^{T}_{i,t+1}
&=100\,d_{it}
\left(b_{i,t+1}-b^{Transitory}_{i,t+1}\right),\\
Y^{P}_{i,t+1}
&=100\,d_{it}
\left(b_{i,t+1}-b^{Persistent}_{i,t+1}\right),
\end{aligned}
\end{equation}
where $d_{it}=\operatorname{sgn}(0.5-b_{i,t-1})$ orients both deviations in the direction from the transitory to the persistent prediction. Positive values of $Y^{T}_{i,t+1}$ indicate departures from the transitory benchmark toward the persistent prediction; negative values of $Y^{P}_{i,t+1}$ indicate departures from the persistent benchmark toward the transitory prediction. Both measures are expressed in percentage points, with zero indicating agreement with the respective benchmark. We average $Y^{T}_{i,t+1}$ and $Y^{P}_{i,t+1}$ within participants and then compute treatment means, giving equal weight to each participant with at least one reset immediately followed by a diagnostic signal.\\

\begin{figure}[H]
    \centering
    \includegraphics[width=0.8\linewidth]{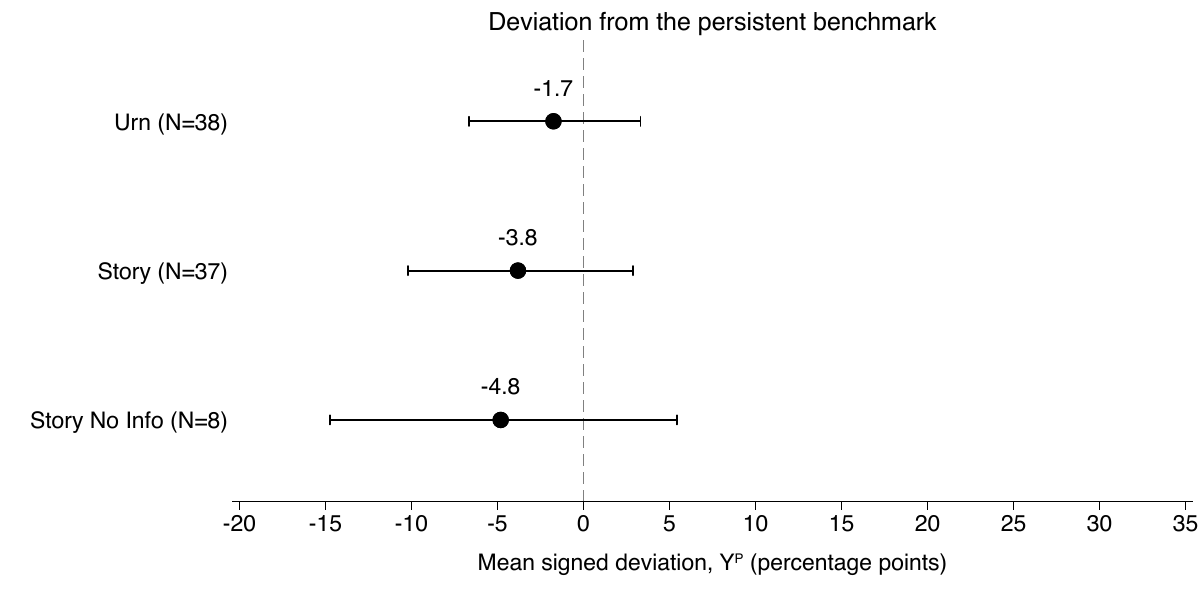}
    \caption{Deviations from the persistent benchmark.\\
    {\footnotesize\textit{Note:} The figure reports mean signed deviations from the Bayesian prediction based on the reset belief of $0.5$. Negative values indicate deviations toward the transitory benchmark. Dots show treatment means of participant-level averages; whiskers show 95\% percentile confidence intervals from $10{,}000$ participant-level bootstrap replications. $N$ denotes participants with at least one qualifying episode.}}
    \label{fig:persistence}
\end{figure}

\begin{figure}[H]
    \centering
    \includegraphics[width=0.8\linewidth]{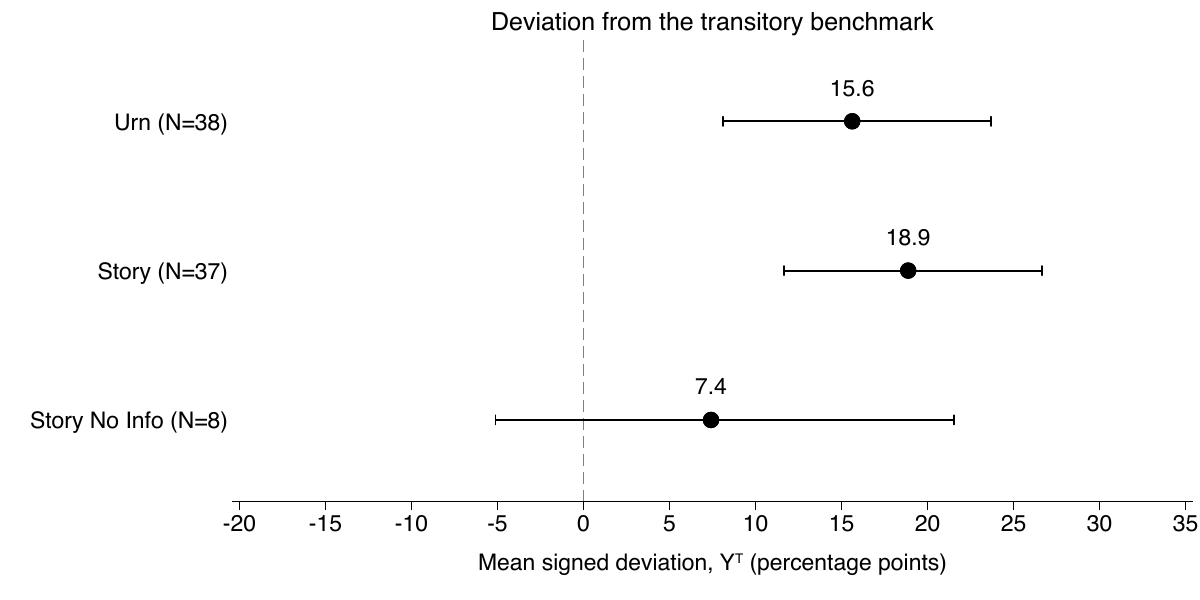}
    \caption{Deviations from the transitory benchmark.\\
    {\footnotesize\textit{Note:} The figure reports mean signed deviations from the Bayesian prediction based on the pre-reset belief. Positive values indicate deviations toward the persistent benchmark. Dots show treatment means of participant-level averages; whiskers show 95\% percentile confidence intervals from $10{,}000$ participant-level bootstrap replications. $N$ denotes participants with at least one qualifying episode.}}
    \label{fig:transitory}
\end{figure}

Figures~\ref{fig:persistence} and~\ref{fig:transitory} report treatment-specific mean signed deviations from the persistent and transitory benchmarks, respectively, with 95\% percentile confidence intervals from $10{,}000$ participant-level bootstrap replications conducted separately within each treatment.

The figures show that subsequent belief reports depart systematically from the transitory benchmark in both \Control and \Narrative. Mean signed deviations are $15.6$ and $18.9$ percentage points, respectively, and both 95\% confidence intervals lie entirely above zero. These departures are in the direction of the persistent prediction, providing evidence against participants simply reinstating their pre-reset beliefs and updating them according to Bayes' rule. By contrast, mean deviations from the persistent benchmark are smaller in absolute magnitude, at $-1.7$ percentage points in \Control and $-3.8$ in \Narrative, and neither confidence interval excludes zero. Thus, average signed deviations are consistent with the persistent benchmark. In \NarrativeNI, the estimates are too imprecise to distinguish between the two benchmarks, reflecting the small number of participants with qualifying resets.\\

Overall, our analysis shows that nondiagnostic information can induce exact belief resets to $0.5$, particularly in \Narrative, whereas such resets are almost absent in \NarrativeNI. The analysis of subsequent diagnostic updates provides evidence against these resets being merely transitory. in both \Control and \Narrative, mean signed deviations from the transitory benchmark are significantly positive, while those from the persistent benchmark are not significantly different from zero. These findings are consistent with persistence, although they do not establish that participants actually use $0.5$ as their prior.

\subsection{Behavioral Types}\label{BTs}

The aggregate frequency of belief resets does not reveal whether they reflect occasional deviations across participants or a systematic updating pattern among a substantial subset of them. To investigate this distinction, we estimate individual-level belief-updating rules that differ in their responses to nondiagnostic signals: leaving beliefs unchanged or resetting them to $0.5$.\\

Our main interest is in how participants respond to nondiagnostic signals, rather than in characterizing biases in diagnostic updating. However, since the classification uses all updating rounds, differences in responses to diagnostic signals may also affect which rule best describes a participant. Our pre-registered analysis documents systematic overreaction to diagnostic signals (Appendix~\ref{app:pre-reg}). Imposing strictly Bayesian updating could therefore lead to misclassification of participants who reset their beliefs after non-diagnostic signals but react differently to diagnostic evidence. We address this by allowing a subject-specific parameter governing the weight assigned to diagnostic signals, reducing the risk that heterogeneity in diagnostic updating drives the classification.\\

Our primary comparison is between two updating rules. The first, which we call \emph{Bayes} for brevity, follows a Bayesian-style updating rule with a subject-specific parameter capturing underreaction or overreaction to diagnostic signals. It leaves beliefs unchanged after nondiagnostic signals. The second, \emph{Bayes-Reset}, follows the same rule for diagnostic signals but resets beliefs to $0.5$ after nondiagnostic ones. The two rules therefore differ only in their responses to nondiagnostic information.\\

We also include two auxiliary rules, \emph{Coarse} and \emph{Coarse-Reset}, to account for participants whose reports may reflect simple reactions to the most recent signal rather than probabilistic belief updating. Both assign extreme beliefs ($0$ or $1$) following diagnostic signals. \emph{Coarse} leaves beliefs unchanged after nondiagnostic signals, whereas \emph{Coarse-Reset} assigns $0.5$. These rules allow us to rule out an alternative explanation for the reset behavior documented in the preceding sections: if a substantial share of participants were classified under \emph{Coarse-Reset}, apparent resets could reflect mechanical, signal-contingent responses rather than departures from an otherwise probabilistic updating process. The four rules are motivated by the patterns observed in the data and are not intended to provide an exhaustive classification of belief-updating behavior.\footnote{See Appendix~\ref{app:types} for a detailed description of the updating rules, estimation procedure, and supplementary results.}\\

Formally, let $b_{t-1} \in [0,1]$ denote the subject's prior belief held at the start of round $t$. Let $s_t \in \{-1,0,+1\}$ denote the realized signal, where $s_t=+1$ favors the true state, $s_t=-1$ disfavors it, and $s_t=0$ is nondiagnostic. For each of the four candidate rules $k$, we define a rule-specific predicted belief $m_t^k$. Reported beliefs are modeled as noisy realizations around this prediction, with support restricted to the feasible belief interval:
\[
b_t \mid k \sim \mathcal{N}_{[0,1]}\left(m_t^k,\sigma_k^2\right),
\]
where \(\mathcal{N}_{[0,1]}(m_t^k,\sigma_k^2)\) denotes a normal distribution with location parameter \(m_t^k\) and scale parameter \(\sigma_k\), truncated to the interval \([0,1]\). In estimation, we restrict the scale parameter to \(\sigma_k\in(0,0.5]\). Each type is therefore characterized by a deterministic updating rule and a rule-specific noise parameter.\\

For each subject, we evaluate the fit of all four candidate rules using the complete set of available updating rounds. The \emph{Bayes} and \emph{Bayes-Reset} rules allow for over- or under-reaction to diagnostic signals and therefore include an additional free parameter relative to the two auxiliary rules. To account for differences in model complexity, we compare the rules using the Bayesian Information Criterion (BIC), which penalizes specifications with additional parameters. Each subject is classified according to the rule with the lowest BIC. This classification allows us to examine whether \Narrative shifts the distribution of updating rules, particularly toward \emph{Bayes-Reset}, which combines probabilistic updating following diagnostic signals with a reset to $0.5$ after nondiagnostic signals.\\

To assess sampling uncertainty in the estimated shares of updating rules, we implement a subject-level nonparametric bootstrap with $10,000$ replications. In each replication, subjects are resampled with replacement separately within each experimental condition, retaining their complete updating histories to preserve within-subject dependence. We then repeat the estimation and BIC-based classification procedure and recompute the condition-specific shares of subjects assigned to each rule. Table~\ref{tab:shares_types} reports the estimated shares for each updating rule and experimental condition.

\begin{table}[H]
\centering\small
\caption{Shares of behavioral types.}\label{tab:shares_types}
\begin{tabular}{lcccc}
\toprule
Treatment
& Bayes
& Bayes-Reset
& Coarse
& Coarse-Reset\\
\midrule
\Control
& \makecell{0.655\\{}[0.576, 0.734]}
& \makecell{0.216\\{}[0.151, 0.288]}
& \makecell{0.022\\{}[0.000, 0.050]}
& \makecell{0.108\\{}[0.058, 0.158]}
\\
\addlinespace[0.5em]
\Narrative
& \makecell{0.469\\{}[0.383, 0.555]}
& \makecell{0.367 \\{}[0.289, 0.453]}
& \makecell{0.063 \\{}[0.023, 0.109]}
& \makecell{0.102 \\{}[0.055, 0.156]}
\\
\addlinespace[0.5em]
\NarrativeNI
& \makecell{0.830 \\{}[0.759, 0.893]}
& \makecell{0.089 \\{}[0.045, 0.143]}
& \makecell{0.054 \\{}[0.018, 0.098]}
& \makecell{0.027 \\{}[0.000, 0.063]}
\\
\midrule
Difference
& & &  & \\
\midrule
\Narrative $-$ \Control
& \makecell{\textbf{-0.187}\\{}[-0.301, -0.065]\\$(0.002)$}
& \makecell{\textbf{0.152}\\{}[0.040, 0.258]\\$( 0.007)$}
& \makecell{0.040\\{}[-0.005, 0.088]\\$(0.087)$}
& \makecell{-0.006\\{}[-0.080, 0.065]\\$(0.895)$}\\
\addlinespace[0.5em]
\Narrative $-$ \NarrativeNI
& \makecell{\textbf{-0.361}\\{}[-0.469, -0.248]\\$(<0.001)$}
& \makecell{\textbf{0.278}\\{}[0.179, 0.373]\\$(<0.001)$}
& \makecell{0.009\\{}[-0.050, 0.067]\\$(0.784)$}
& \makecell{\textbf{0.075}\\{}[0.017, 0.137]\\$(0.011)$}\\
\bottomrule
\end{tabular}
\caption*{\footnotesize\textit{Note:} The table reports the median share of subjects classified under each behavioral rule along with the corresponding 95\% confidence intervals in square brackets. The bottom panel reports the difference in estimated shares between treatments. The corresponding 95\% confidence intervals are reported in square brackets, with p-values reported below. Two-sided bootstrap $p$-values for differences between treatments are reported in parentheses. Treatment differences that are statistically significant at the 95\% level are shown in bold text.}
\end{table}

Table~\ref{tab:shares_types} shows that 36.7\% of participants in \Narrative are classified under \textit{Bayes-Reset}, compared with 21.6\% in \Control and 8.9\% in \NarrativeNI. Both differences are statistically significant. Conversely, the share classified under \textit{Bayes} is 46.9\% in \Narrative, compared with 65.5\% in \Control and 83.0\% in \NarrativeNI, with both differences also statistically significant.\\

Differences in the shares classified under \textit{Coarse} are not statistically significant. For \textit{Coarse-Reset}, the share is significantly lower in \NarrativeNI (2.7\%) than in \Narrative (10.2\%), while the difference between \Control (10.8\%) and \Narrative is not statistically significant.\\

Overall, the classification suggests that resets to $0.5$ are a systematic feature of the updating rule that best describes a substantial share of participants in \Narrative, rather than simply reflecting mechanical responses to the most recent signal.

\vspace{0.5cm}
\section{Discussion}\label{sec:discussion}

A distinctive feature of our findings is that nondiagnostic signals can induce revisions to exactly $0.5$, even from previously reported beliefs far from this value. The question is therefore why participants report exactly $0.5$, rather than simply revise their beliefs toward it. We consider explanations related to the reporting process, such as response noise and anchoring, as well as mechanisms involving uncertainty about the appropriate probability assessment. Such uncertainty may be particularly relevant in a narrative setting, where clues may invite speculation about alternative explanations, despite the explicitly specified signal-generating process.\\

Reporting noise can generate apparent belief revisions even when underlying beliefs remain unchanged. For example, a previous report may be unusually extreme partly because of a reporting error. If that error is not repeated, the next report may be less extreme, creating the appearance of a revision toward $0.5$ after a nondiagnostic signal. This regression-to-the-mean effect, however, predicts movement toward the underlying belief, which need not equal $0.5$, and does not by itself explain a concentration of revisions at exactly $0.5$.
Accounting for this pattern and its variation across treatments requires an explanation of how underlying assessments and reporting processes produce $0.5$ responses that random noise alone cannot account for.\\
Explanations based on reference points are also insufficient on their own. The elicitation slider displayed no default response, ruling out mechanically generated focal points. We also find little indication that participants return to their initially reported prior.\footnote{Figure~\ref{fig:bubble_return_prior} in the appendix plots reported beliefs after receiving nondiagnostic information against initial priors, excluding observations where the preceding belief is equal to $0.5$. A return to the initial prior would produce a concentration of observations around the diagonal, which we do not observe.} Moreover, $0.5$ reports are nearly absent in \NarrativeNI, suggesting that the salience of $0.5$ alone cannot explain the treatment differences. These observations motivate considering mechanisms that can make reporting exactly 0.5 attractive, rather than simply producing less extreme reports.\\

We next consider cognitive uncertainty about the appropriate probability assessment. \citet{EnkeGraeber2023} show that greater cognitive uncertainty is associated with reports compressed toward $0.5$. In our setting, nondiagnostic narrative clues may make participants less certain about how to interpret the accumulated evidence, even though the signal probabilities are explicitly specified. The virtual absence of exact resets in \NarrativeNI does not rule out this explanation: an explicit no-information message may make it easier to retain the previous assessment, whereas an additional clue may prompt participants to reconsider it. However, an explanation based on cognitive uncertainty must account for why this reconsideration produces revisions to exactly $0.5$, rather than merely less extreme reports. Because we do not directly measure cognitive uncertainty, we cannot determine its role in generating the observed resets.\\


The last mechanism we consider is ambiguity aversion. Although we do not measure it directly, it could explain why many participants report the two states as equally likely, rather than merely shifting their reports towards $0.5$. In multiple-priors models with \textit{maxmin} preferences \citep{GilboaSchmeidler1989,GhirardatoMaccheroniMarinacci2004}, reporting $0.5$ maximizes the worst-case expected payoff whenever the set of plausible probabilities includes $0.5$. This property holds for symmetric strictly proper scoring rules under linear utility over money, including the quadratic rule used in our experiment \citep{Chambers2008,ChambersHealyLambert2019}. Importantly, this mechanism can produce $0.5$ reports across a range of plausible probability assessments, rather than merely shifting reports towards $0.5$. If greater perceived ambiguity expands the set of plausible probabilities to include $0.5$, such reports may become more frequent. 

\vspace{0.5cm}
\section{Economic Consequences}\label{sec:simulation}



Our experiment documents belief resets following nondiagnostic signals, but whether these distortions have meaningful economic consequences is a separate question. Comparing experimental earnings under a quadratic scoring rule is of limited interest in this respect, as economic payoffs typically depend on the actions people take and their timing, rather than on reported probabilities. One potential consequence is that belief resets may delay payoff-relevant actions. We therefore simulate a sequential decision problem to examine whether differences in reset behavior across treatments affect the number of signals required to reach a decision threshold. We also examine their implications for belief accuracy over longer horizons, measured by the excess Brier score relative to a Bayesian benchmark. This exercise illustrates potential economic consequences rather than providing estimates of real-world decision costs.\\

We conduct a Montecarlo simulation in which agents begin with a prior belief of 0.5 and subsequently observe signals generated by the same stochastic process as in the experiment. Diagnostic signals are incorporated using Bayes' rule.\footnote{We hold diagnostic updating fixed across treatments to isolate the consequences of differences in reset behavior, rather than reproducing all observed deviations from Bayesian updating.} Following a nondiagnostic signal, beliefs reset to 0.5 with a treatment-specific probability estimated from the experimental data. This probability depends on the absolute distance of the preceding belief from 0.5, with linear interpolation between estimated points. In the absence of a reset, beliefs remain unchanged.\footnote{The treatment-specific reset probabilities are based on the probit estimates reported in Table~\ref{reg:prob} and illustrated in Figure~\ref{fig:prob_midpoint}.}\\

Each simulated belief path continues until the posterior crosses one of three symmetric decision thresholds: (0.20, 0.80), (0.10, 0.90), or (0.05, 0.95). These thresholds capture environments in which agents require some degree of confidence before taking an action. Therefore, the outcome of interest is represented by the stopping time: the number of signals required to reach a decision threshold. To capture the policy-relevant margin identified by the experiment---the delay induced by exposure to nondiagnostic narrative clues---we compare \Narrative with \NarrativeNI. In many settings, narratives cannot realistically be removed, but institutions can decide whether to show, filter, or withhold information deemed irrelevant. We also report results for the \Control condition and a simulated Bayesian agent as benchmarks.\\

Additionally, we assess belief accuracy using the excess Brier score relative to a Bayesian benchmark. For each updating rule, we calculate the average difference in squared prediction errors over the first 10, 50, and 100 signals, using the same simulated signal sequences. In this exercise, belief updating continues over the entire evaluation horizon, independently of whether a decision threshold has been crossed.\\

The simulation results, reported in Table~\ref{tab:stopping_times}, show that midpoint reversion substantially slows belief convergence. Panel A reports the stopping times under alternative updating rules. The median stopping time is larger in \Narrative compared to \NarrativeNI at all decision thresholds. At the first threshold (0.20, 0.80), it rises from 14 rounds in \NarrativeNI to 17 rounds in \Narrative (+21\%). At the intermediate threshold (0.10, 0.90), the median stopping time increases from 28 to 39 rounds (+39\%). The effect is larger under the stricter threshold (0.05, 0.95), where the median stopping time rises from 42 to 74 rounds (+76\%).\\

Panel B shows that midpoint reversion also reduces belief accuracy. Over the first 10 signals, the average excess Brier score relative to Bayesian updating is 0.006 in both \Narrative and \Control, compared with 0.001 in \NarrativeNI. The differences become more pronounced over longer horizons. At $T=50$, the excess Brier score is 0.050 in \Narrative, compared with 0.039 in \Control and 0.005 in \NarrativeNI. At $T=100$, these values increase to 0.088, 0.069, and 0.008, respectively. Thus, under the simulated updating rules, the loss of belief accuracy becomes more pronounced over longer horizons.

\begin{table}[H]
\centering\small
\caption{Stopping times and accuracy under alternative updating rules.}

\textit{Panel A: Stopping times}

\begin{tabular*}{0.85\linewidth}{@{\extracolsep{\fill}}llcc@{}}
\toprule
& & \multicolumn{2}{c}{Number of signals} \\
\cmidrule(lr){3-4}
Threshold & Update rule & Mean & Median \\
\midrule
\multirow{4}{*}{ $(0.20, 0.80)$ }
& Bayesian                 & 17.91 & 14 \\
& Reset--\Control          & 21.92 & 17 \\
& Reset--\Narrative        & 22.37 & 17 \\
& Reset--\NarrativeNI      & 18.67 & 14 \\
\addlinespace

\multirow{4}{*}{ $(0.10, 0.90)$ }
& Bayesian                 & 33.57 & 26 \\
& Reset--\Control          & 47.89 & 36 \\
& Reset--\Narrative        & 52.62 & 39 \\
& Reset--\NarrativeNI      & 35.43 & 28 \\
\addlinespace

\multirow{4}{*}{ $(0.05, 0.95)$ }
& Bayesian                 & 49.42 & 40 \\
& Reset--\Control          & 81.92 & 61 \\
& Reset--\Narrative        & 100.56 & 74 \\
& Reset--\NarrativeNI      & 52.40 & 42 \\
\bottomrule
\end{tabular*}

\vspace{1em}

\textit{Panel B: Belief accuracy}

\begin{tabular*}{0.85\linewidth}{@{\extracolsep{\fill}}lccc@{}}
\toprule
& \multicolumn{3}{c}{Average excess Brier score} \\
\cmidrule(lr){2-4}
Update rule & $T=10$ & $T=50$ & $T=100$ \\
\midrule
Bayesian                 & 0.000 & 0.000 & 0.000 \\
Reset--\Control          & 0.006 & 0.039 & 0.069 \\
Reset--\Narrative        & 0.006 & 0.050 & 0.088 \\
Reset--\NarrativeNI      & 0.001 & 0.005 & 0.008 \\
\bottomrule
\end{tabular*}

\caption*{\footnotesize \textit{Note}: Panel A reports the mean and median number of signals required for beliefs to cross each decision threshold. Panel B reports average excess Brier scores relative to Bayesian updating over the first $T=10$, $50$, and $100$ signals, respectively. Excess Brier scores are calculated as the mean difference in squared prediction errors between each updating rule and the Bayesian benchmark, averaged across simulated signal sequences and periods $t=1,\ldots,T$. All updating rules are evaluated on the same signal sequences, and belief updating continues after any decision threshold is crossed. The excess Brier score is zero by construction for Bayesian updating.}
\label{tab:stopping_times}
\end{table}

The simulation shows that midpoint reversion results in longer stopping times and lower belief accuracy. The excess Brier scores indicate that the loss of accuracy relative to the Bayesian benchmark becomes more pronounced over longer simulated horizons, particularly in \Narrative compared with \NarrativeNI. By delaying threshold crossing, nondiagnostic narrative clues increase the amount of evidence agents must process before taking action. This delay is economically relevant whenever information acquisition is costly, or when delay itself reduces payoffs because opportunities expire, competitors move first, or decisions become less valuable over time. The implication is therefore not only that nondiagnostic narrative clues can reduce belief accuracy, but also that, under the simulated updating rules, they can defer action in environments where timely decisions are valuable.


\vspace{0.5cm}
\section{Conclusions}\label{sec:conclusions}

This paper provides experimental evidence that nondiagnostic details embedded in a narrative distort reported belief updating. In a controlled laboratory experiment, we hold the signal-generating process constant across three treatments: a fictional investigation story with diagnostic and nondiagnostic clues, the same story with nondiagnostic clues replaced by explicit no-information messages, and an abstract urn problem. In all treatments, participants repeatedly report incentivized beliefs about which of two mutually exclusive states is true.\\

We find a systematic departure from Bayesian updating. Nondiagnostic signals induce exact resets of reported beliefs to 0.5, although Bayes’ rule prescribes no revision. This pattern is particularly pronounced when nondiagnostic signals are presented as narrative clues, weaker in the abstract urn problem, and almost absent when those clues are replaced by explicit no-information messages within the same narrative.\\


While our design cleanly identifies the effect of nondiagnostic narrative clues on reported beliefs, it does not pin down the underlying mechanism. One possibility is that narrative clues induce perceived ambiguity about the completeness of the experimental model, contributing to the observed concentration of reports at exactly $0.5$. Distinguishing this interpretation from other possible mechanisms remains a question for future research.



\subsection*{Data and Code Availability}

The replication package for this paper is publicly available at

\url{https://github.com/paolopin/narrative-belief-updating}.

The repository contains the Stata code for the empirical analysis, the MATLAB code for the behavioral-type classification, and the Python code for the numerical simulations.

\bibliography{literature}

\newpage

\appendix

\pagenumbering{arabic}
\setcounter{footnote}{0}

\addtocontents{toc}{\protect\setcounter{tocdepth}{2}}

\setcounter{table}{0}
\renewcommand{\thetable}{\thesection.\arabic{table}}
\counterwithin{table}{section}
\renewcommand\thefigure{\thesection.\arabic{figure}}
\counterwithin{figure}{section}

\section*{Online Appendix for ``Fooling Yourself: how narratives shape beliefs''}

\subsubsection*{A. Albertazzi, P. Pin, M. Stimolo and A. Stringhi}

\subsubsection*{For Online Publication}

\vspace{2cm}

\tableofcontents

\newpage

\section{Experimental Instructions}\label{instructions}
\subsection*{Consent Form}

This research is conducted by the University of Siena and IMT School for Advanced Studies Lucca. Before deciding whether to participate in the research, we invite you to carefully read this page.

\paragraph{Voluntary Nature of Participation}

Participation in this study is voluntary and does not involve any benefits related to your university career, nor any consequences in case of non-participation. Should you choose to participate, you may withdraw at any time without the need to provide any explanation. Leaving the activity will not result in any penalties other than the loss of any money accumulated during the session. The expected duration of this session is approximately 30 minutes.

\paragraph{Procedure}

The activity consists of several parts and a demographic information questionnaire. You will receive instructions for each part upon completion of the previous one. In some parts, you will be required to make decisions. The described procedures will be faithfully implemented.

\paragraph{Compensation for Participation}

For your participation, you will receive 4 Euros plus an additional contribution ranging from 0 to 11 Euros. The additional contribution will be expressed in points and converted at the end of the study at the following exchange rate: 100 points for 1 Euro. There will be no costs to participate in this activity, other than the time the participant chooses to dedicate to it.

\paragraph{Payments}

Payments will be made via Amazon gift cards by Business Intelligence Group S.r.l., which collaborates with the University of Siena. To receive payment, you will need to provide an email address where you would like to receive the Amazon gift card at the end of the experiment. Payment will be made within 15 working days.

\paragraph{Data Processing}

All collected data will be stored in a pseudonymous and secure manner on a computer at the University of Siena. They will be used exclusively for the advancement of knowledge and may be made public only through communications and/or scientific publications. The data related to your email address will be used solely for the purpose of processing the payment. Pseudonymous data will always be analyzed in an aggregated form and never on an individual level. The subject to whom the personal data refers has the rights outlined in sections 2, 3, and 4 of Chapter III of the EU GDPR Regulation No. 2016/679 (\url{https://www.garanteprivacy.it/regolamentoue}).\\

\noindent If you experience any discomfort at any time during the session or after its completion, or if you need any clarifications, you may contact the research supervisor, Prof. Paolo Pin, via email at \href{mailto:paolo.pin@unisi.it}{paolo.pin@unisi.it}.\\

\noindent Thank you for reading this information and considering your participation in the study. By continuing to read, you confirm that you have understood the information and declare your intention to participate.

\vspace{1cm}

\subsection*{Instructions Part 1}

\subsubsection*{[ \Control]}

In this section, you will only be asked to carefully read the following text.\\

\noindent \uline{Please read the text below carefully:}\\

An experiment was conducted to determine from which of two urns a series of balls was drawn. The experimenters suspect that the chosen urn is either the one containing mostly \textcolor{red}{\textbf{Color 1}} balls or the one containing mostly \textcolor{blue}{\textbf{Color 2}} balls, but they cannot know for sure. Three balls were randomly drawn from the urn in sequence. The first ball was \textcolor{red}{\textbf{Color 1}}, followed by a second ball of \textcolor{blue}{\textbf{Color 2}}. Finally, the third ball drawn was \textbf{White}.

\subsubsection*{[ \Narrative \& \NarrativeNI]}

In this section, you will only be asked to carefully read the following text.\\

\noindent \uline{Please read the text below carefully:}\\

A small town has been shaken by a series of thefts. Locals suspect that the culprit could be either \textbf{Suspect 1}, a quiet and reserved man, or \textbf{Suspect 2}, known for his outgoing personality, but they have no way to prove it. In the past, \textbf{Suspect 1} underwent medical treatment for kleptomania. \textbf{Suspect 2} had a minor criminal record for petty thefts a few years ago. A local artist created sketches of both suspects based on eyewitness descriptions, but the portraits looked very much alike, making the identification process even more difficult

\vspace{1cm}

\subsection*{Instructions – Part 2}

In this part, we will ask you to indicate, based on the information you received in Part 1, which of the two alternatives you consider to be more likely.\\

\noindent To express your judgment, you will use a slider like the one shown at the bottom of the page. The more likely you believe one of the two options is, the closer you should move the slider toward it.    
You will be paid based on how accurate your response is. The closer your answer is to the correct option, the higher your payment. The farther your answer is from the correct option, the lower your payment.\\

\noindent \uline{Simply put, it is in your best interest to choose the slider value that you believe is most accurate.}\\

\noindent For clarity, we will display your expected payment on the screen as you make your decision, using the following formula:

\[100\times[1-(v-r)^2]\]

\noindent where $v$ is the true value (0 if the correct option is \textbf{Option 1}, or 1 if the correct option is \textbf{Option 2}), and $r$ is the response you provide.\\

\noindent Please note that the payment is expressed in points and \uline{100 points correspond to 1 euro}.\\

\noindent Below, you can see an example of the slider and how your bonus payment is calculated.
To select your preferred value, simply click on the slider until it reaches the desired number. We invite you to practice using the slider below and to pay attention to how the expected payments change.\\

\[[SLIDER]\]

\subsection*{Comprehension Questions}
5 questions with examples.

\subsection*{Prior Elicitation}

\uline{Now we will show you again the scenario you read in Part 1.}

\[[TEXT]\]

\noindent \uline{Adjust the slider to indicate which of the two alternatives you believe is more likely.} We ask you to indicate your answer using the slider. Please remember that it is in your best interest to provide the answer you believe to be most accurate.\\

\[[SLIDER]\]

\vspace{1cm}
\subsection*{Instructions Part 3}

\subsubsection*{[ \Control]}

In this part, you will receive a series of additional pieces of information related to the situation presented so far. After receiving each new piece of information, you will again be asked to indicate which option you believe is more likely, using a slider just like in the previous section. The slider will function in the same way, and the expected payment will be calculated using the same method. Each of your answers will be rewarded, and we remind you that \uline{100 points correspond to 1 euro}.\\

\noindent \uline{The new information you will receive can be of three types:}\\

\begin{itemize}
    \item A ball of \textcolor{red}{\textbf{Color 1}} is drawn;
    \item A ball of \textcolor{blue}{\textbf{Color 2}} is drawn;
    \item A ball of color \textbf{White} is drawn.
\end{itemize}

\noindent The new pieces of information you will receive are selected at random, but \uline{the probability with which you see them depends on which of the two alternatives is actually true}. This means that some pieces of information will occur more frequently than others, depending on which alternative is correct.\\

\noindent \uline{The probability with which this information will appear depends on which of the two alternatives is the true one:}

\begin{itemize}
    \item If the urn with a prevalence of \textcolor{red}{\textbf{Color 1}} balls is the one used for the drawing, the probability of drawing a \textcolor{red}{\textbf{Color 1}} ball will be \textbf{45\%}, while the probability of drawing a \textcolor{blue}{\textbf{Color 2}} ball will be \textbf{30\%}. Finally, the probability of drawing a \textbf{White} ball will be \textbf{25\%}.
    \item If the urn with a prevalence of \textcolor{blue}{\textbf{Color 2}} balls is the one used for the drawing, the probability of drawing a \textcolor{blue}{\textbf{Color 2}} ball will be \textbf{45\%}, while the probability of drawing a \textcolor{red}{\textbf{Color 1}} ball will be \textbf{30\%}. Finally, the probability of drawing a \textbf{White} ball will be \textbf{25\%}.
\end{itemize}

\vspace{1cm}

\subsubsection*{[ \Narrative]}

In this part, you will receive a series of additional pieces of information related to the situation presented so far. After receiving each new piece of information, you will again be asked to indicate which option you believe is more likely, using a slider just like in the previous section. The slider will function in the same way, and the expected payment will be calculated using the same method. Each of your answers will be rewarded, and we remind you that \uline{100 points correspond to 1 euro}.
\\

\noindent \uline{The new information you will receive can be of three types:}\\

\begin{itemize}
    \item Incriminating information for \textbf{Suspect 1};
    \item Incriminating information for \textbf{Suspect 2};
    \item Non-incriminating information (no suspect is mentioned by name).
\end{itemize}

\noindent The new pieces of information you will receive are selected at random, but \uline{the probability with which you see them depends on which of the two alternatives is actually true}. This means that some pieces of information will occur more frequently than others, depending on which alternative is correct.\\

\noindent \uline{The probability with which this information will appear depends on which of the two alternatives is the true one:}

\begin{itemize}
    \item If \textbf{Suspect 1} is the true culprit, the probability of receiving information incriminating them will be \textbf{45\%}, while for \textbf{Suspect 2} it will be \textbf{30\%}. Finally, the probability of receiving non-incriminating information will be \textbf{25\%}.
    \item If \textbf{Suspect 2} is the true culprit, the probability of receiving information incriminating them will be \textbf{45\%}, while for \textbf{Suspect 1} it will be \textbf{30\%}. Finally, the probability of receiving non-incriminating information will be \textbf{25\%}.
\end{itemize}

\vspace{1cm}

\subsubsection*{[ \NarrativeNI]}
In this part, you will receive a series of additional pieces of information related to the situation presented so far. With a 25\% probability you will not receive any information.
After receiving (or not) each new piece of information, you will again be asked to indicate which option you believe is more likely, using a slider just like in the previous section. The slider will function in the same way, and the expected payment will be calculated using the same method. Each of your answers will be rewarded, and we remind you that \uline{100 points correspond to 1 euro}.
\\

\noindent \uline{The new information you will receive can be of two types:}\\

\begin{itemize}
    \item Incriminating information for \textbf{Suspect 1};
    \item Incriminating information for \textbf{Suspect 2}.
\end{itemize}

\noindent \textbf{However, every new piece of information has a probability of being withheld.}\\

\noindent The new pieces of information you might receive are selected at random, but \uline{the probability with which you see them depends on which of the two alternatives is actually true}. This means that some pieces of information will occur more frequently than others, depending on which alternative is correct.\\

\noindent \uline{The probability with which this information will appear depends on which of the two alternatives is the true one:}

\begin{itemize}
    \item If \textbf{Suspect 1} is the true culprit, the probability of receiving information incriminating them will be \textbf{45\%}, while for \textbf{Suspect 2} it will be \textbf{30\%}. Finally, the probability of not receiving any information will be \textbf{25\%}.
    \item If \textbf{Suspect 2} is the true culprit, the probability of receiving information incriminating them will be \textbf{45\%}, while for \textbf{Suspect 1} it will be \textbf{30\%}. Finally, the probability of not receiving any information will be \textbf{25\%}.
\end{itemize}

\vspace{0.5cm}

\subsection*{List of Diagnostic signals}
\begin{enumerate}
\item (Name of suspect) has recently experienced financial difficulties and took out a loan to cover unexpected expenses.
\item (Name of suspect) was seen wearing gloves and a hat on the night of one of the burglaries, which seemed unusual given that the day had been hot.
\item (Name of suspect) bragged about recent expensive purchases despite lacking a steady job.
\item (Name of suspect) was recently going through hard times and struggling to make ends meet.
\item (Name of suspect) was facing financial difficulties and was overheard talking about the value of the stolen items.
\item (Name of suspect) is known to have a gambling problem and owes money to several people in town.
\item (Name of suspect) was seen carrying a large bag into their home on the night of one of the burglaries; the bag appeared heavy and full of objects.
\item (Name of suspect) was seen loitering around several victims’ houses shortly before the burglaries occurred.
\item (Name of suspect) was a known hoarder, and the way they had come to possess certain items seemed suspicious.
\item (Name of suspect) had an unverified alibi for the time of the burglaries and was seen with expensive items they could not explain.
\end{enumerate}
\vspace{0.5cm}

\subsection*{List of Non-diagnostic signals}
\begin{enumerate}
\item One of the stolen items was a valuable antique necklace passed down through generations in a prominent local family.
\item Some of the city’s security cameras malfunctioned during the period of the burglaries, making it difficult to obtain visual evidence.
\item The stolen items included expensive electronics such as smartphones and laptops.
\item The stolen items included a rare collection of vintage coins that had been on display at the city’s historical museum.
\item One of the stolen items—a rare painting—was found abandoned in a hidden shack in a nearby forest.
\item An anonymous letter was sent to the local newspaper claiming responsibility for the burglaries but providing no concrete proof or details.
\item The city’s mayor offered a substantial reward to anyone who could provide information leading to the thief’s arrest, intensifying search efforts.
\item A witness reported seeing a suspicious van parked near one of the crime scenes.
\item A hidden camera was installed in a store that had previously been targeted, but it failed to capture any incriminating footage.
\item The city’s mayor offered a substantial reward to anyone who could provide information leading to the thief’s arrest.
\end{enumerate}

\newpage

\section{Validation of the Story Scenario }\label{validation}

To ensure that the introductory scenario used in the \Narrative and \NarrativeNI conditions was perceived as neutral and symmetric, we conducted a series of pre-registered online validation studies on independent samples. This validation aimed to minimize any implicit biases that the story might induce.\footnote{The pre-registrations can be accessed via the following link: \href{https://osf.io/b5ymf/overview}{https://osf.io/b5ymf/overview}.}\par

We validated both the main event and the associated clues through a multi-step procedure:

\begin{itemize}
\item \textbf{Main Event Selection:} Participants were shown several alternative stories and asked to report their belief (via a slider) about which suspect was more likely to be guilty. Only those stories for which average beliefs were not significantly different from 0.5 were retained. From this subset, one was selected for the main experiment.
\item \textbf{Diagnostic Clue Selection:} Participants were presented with several pairs of directional clues---each pair containing one clue pointing to each suspect---and asked to select the pair they believed would lead another participant to hold beliefs as close as possible to a 50/50 prior. To incentivize accurate judgments, we used data from a previous pilot in which a different group of participants had completed the prior elicitation task after reading these clues. Payment was based on how close the actual prior reported by the earlier participant was to 0.5. This procedure allowed us to identify the pair of clues most consistently judged to be balanced.
\item \textbf{Nondiagnostic Clue Selection:} A similar method was used to validate nondiagnostic clues. Participants were shown the validated main event along with several potential nondiagnostic facts and asked to select the one they believed had led a previous participant to report a belief closest to 0.5. Again, payments were based on the actual priors elicited in the earlier pilot, linking incentives to prediction accuracy. The clues most frequently judged as neutral were selected for use in the main experiment.

\end{itemize}

Importantly, we did not select clues based on the actual distribution of beliefs they generated, as this could result in selecting signals that appear balanced for arbitrary or misleading reasons. Instead, we relied on participants’ perceptions of neutrality, consistent with our goal of controlling how information is expected to be interpreted.\par

All clues retained for the main experiment---both diagnostic and nondiagnostic---were thus validated based on perceived neutrality and symmetry, not on ex-post outcomes. All other clues that were not selected for the introductory scenario were used as signals in Part 3 of the experiment. Since the nature and interpretation of these signals were explained unambiguously to participants, additional validation was not necessary for this purpose.

\newpage
\section{Supplementary Analysis}\label{app:additional_results}

\begin{figure}[H]
    \centering
\includegraphics[width=1\linewidth]{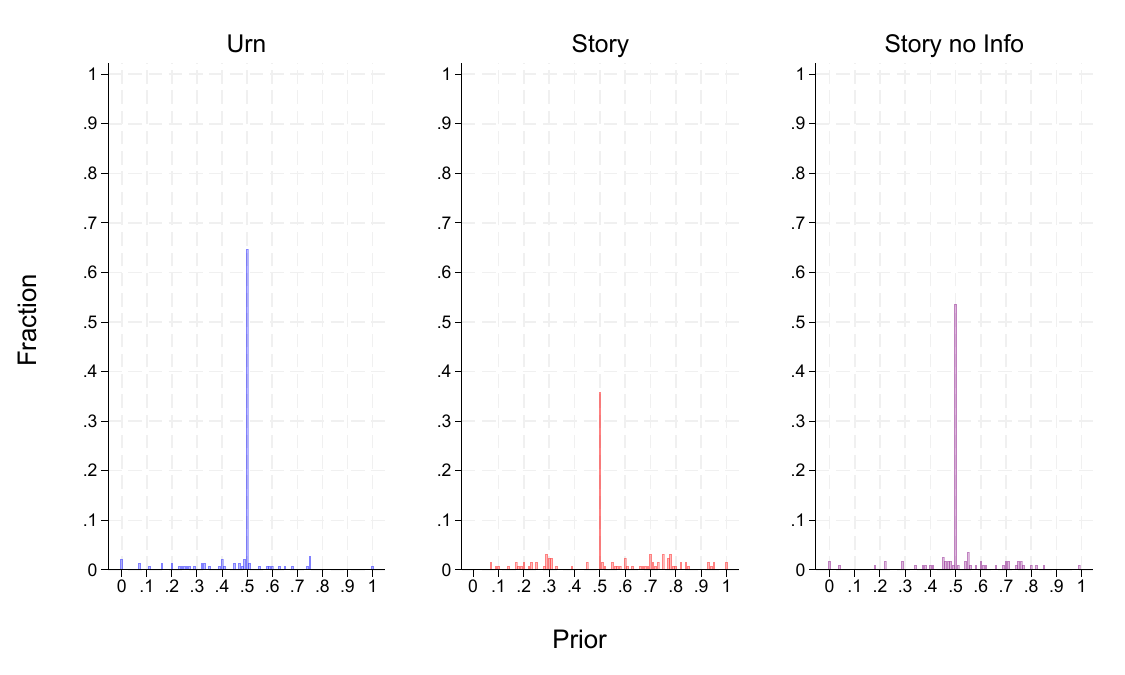}
    \caption{Distribution of initial priors by treatment.}
    \label{fig:priors_distribution}
\end{figure}

Figure \ref{fig:priors_distribution} shows the distribution of beliefs at round 0 for all treatments.

\begin{figure}[H]
    \centering
\includegraphics[width=1\linewidth]{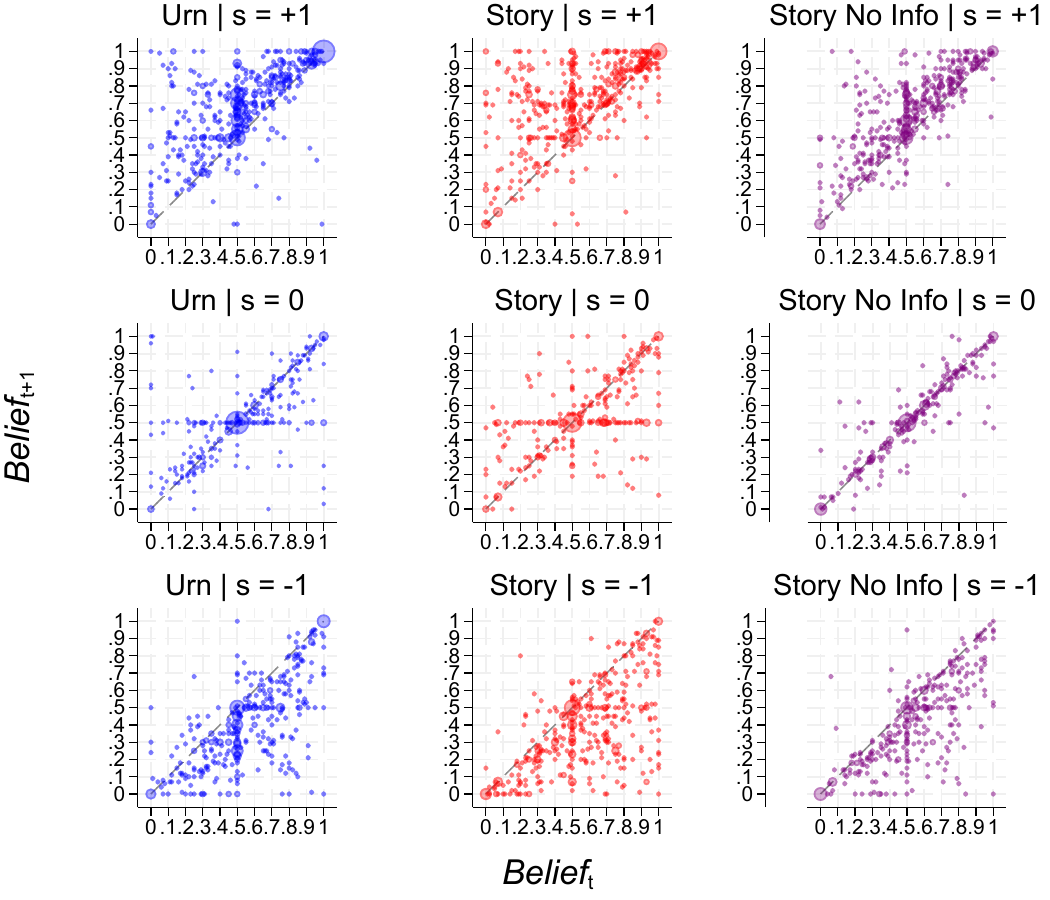}
    \caption{Change in beliefs.} \label{fig:bubbleplots}
    \caption* {\footnotesize\textbf{Notes:} The bubble plots illustrate belief updating: each tile shows how beliefs evolve from round $t$ to round $t+1$, conditional on signal (s = -1, 0, +1) and treatment group (\Control, \Narrative, and \NarrativeNI). The size of the marker indicates the number of data points for that specific coordinate, where the smallest marker refers to one observation only.}
\end{figure}

Figure~\ref{fig:bubbleplots} provides a visual account of how beliefs are updated. Each panel plots participants' stated beliefs in round $t$ (x-axis) against their stated beliefs in round $t+1$ (y-axis), conditional on the realized signal and the experimental condition. Thus,  departures from the 45-degree line (representing \emph{no updating}) reveal both the presence and the direction of belief revisions.

To facilitate interpretation, we recode beliefs so that $1$ always corresponds to assigning probability one to the \emph{true} state. We then code signals as $s\in{-1,0,+1}$, where $s=+1$ ($s=-1$) denotes evidence in favor of the true (false) state and therefore implies an upward (downward) Bayesian update relative to the prior (i.e., the belief stated in the previous round). We set $s=0$ for nondiagnostic signals in \Control and \Narrative, and for the no-information realizations in \NarrativeNI.

\subsection{Robustness checks}

\begin{figure}[H]
    \centering
\includegraphics[width=0.8\linewidth]{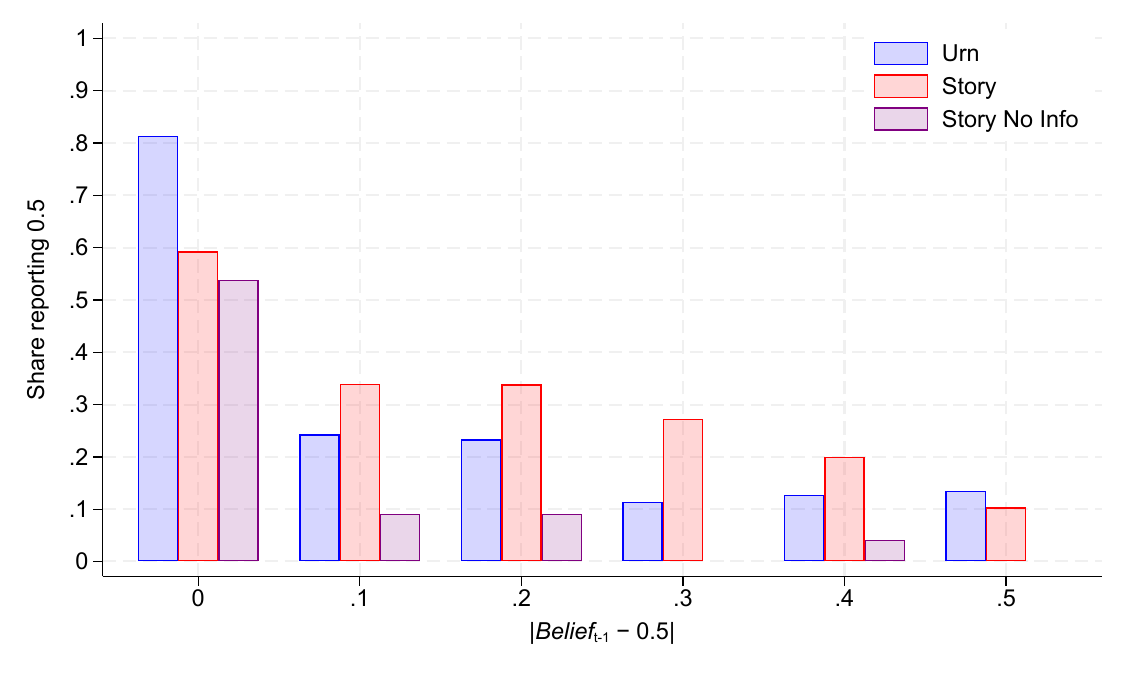}
    \caption{Reporting exactly 0.5 after nondiagnostic signals conditional on belief.\\ {\footnotesize\textit{Note:} The figure shows, for each experimental condition, the share of reports equal to 0.5 following a nondiagnostic signal, conditional on the absolute distance of the prior from 0.5 (binned).}}\label{fig:shares_midpoint}
\end{figure}

Table \ref{reg:prob} reports estimates from the probit specification used in the main analysis, together with an LPM as a robustness check.

\begin{table}[H]\centering
\caption{Probability of reporting 0.5.}\label{reg:prob}
\def\sym#1{\ifmmode^{#1}\else\(^{#1}\)\fi}
	\begin{tabular}{l*{2}{cc}}
		\hline\hline
Pr($Belief = 0.5$)        &\multicolumn{1}{c}{Probit} &\multicolumn{1}{c}{LPM}\\
         \hline
\textit{Urn}       &0.360 (0.45)         &0.055 (0.07)         \\
\textit{Story No Info}      &-1.004\sym{\dagger} (0.56)         &-0.226\sym{**} (0.07)         \\
$d_{t-1}$      &-1.487 (1.07)         &-0.298\sym{\dagger} (0.17)         \\
\textit{Urn}$\times d_{t-1}$ &-2.462 (1.55)         &-0.352 (0.23)         \\
\textit{Story No Info }$\times d_{t-1}$ &-4.714\sym{\dagger} (2.77)         &0.044 (0.23)         \\
Constant    &-0.688 (0.45)         &0.364\sym{***} (0.07)         \\
\hline
Observations&  784         &  784         \\
\hline\hline
\end{tabular}
\caption*{\footnotesize\textit{Note:} Probability of reporting a belief of 0.5 after receiving a nondiagnostic signal. The leftmost column reports estimates from a linear probability model. All models control for round fixed effects, the strength of evidence received up to time $t$, and allow for subject-level intercepts. Standard errors are in parentheses.\footnotesize \sym{\dagger} \(p<0.10\), \sym{*} \(p<0.05\), \sym{**} \(p<0.01\), \sym{***} \(p<0.001\)}
\end{table}

\begin{figure}[H]
    \centering
\includegraphics[width=1\linewidth]{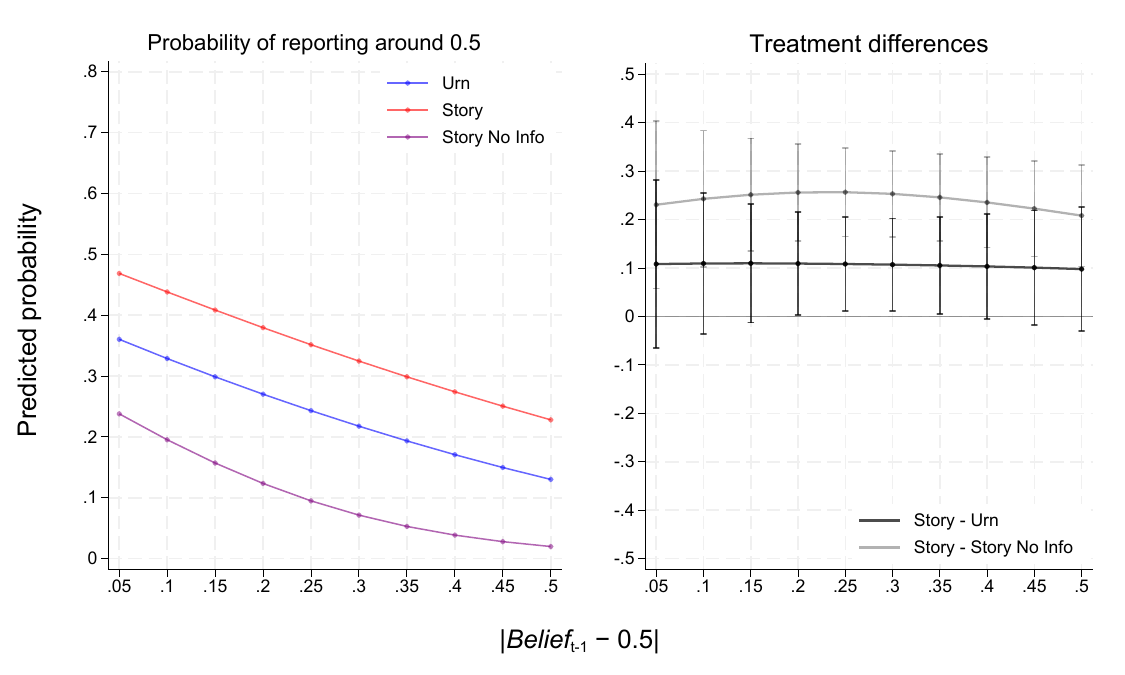}
    \caption{Predicted probability of exact belief reset.\\{\footnotesize\textit{Note:} The left panel shows the predicted probability of reporting a belief in the interval $[0.45, 0.55]$ after observing a nondiagnostic signal for all treatment conditions. The right panel shows the predicted treatment differences for the relevant treatment comparisons. Vertical bars represent 95\% confidence intervals.}}\label{fig:prob_midpoint_robust}
\end{figure}

We then use a broader definition of an exact belief reset by defining an indicator equal to one when the updated belief falls within a neighborhood of $0.5$: $Y_{it}=\mathbf{1}\{b_{it}\in[0.45,0.55]\}$. Figure \ref{fig:prob_midpoint_robust} shows the corresponding predicted probabilities and treatment differences over prior distance from $0.5$.

\subsection{Quadratic specification}

\begin{figure}[H]
    \centering
\includegraphics[width=1\linewidth]{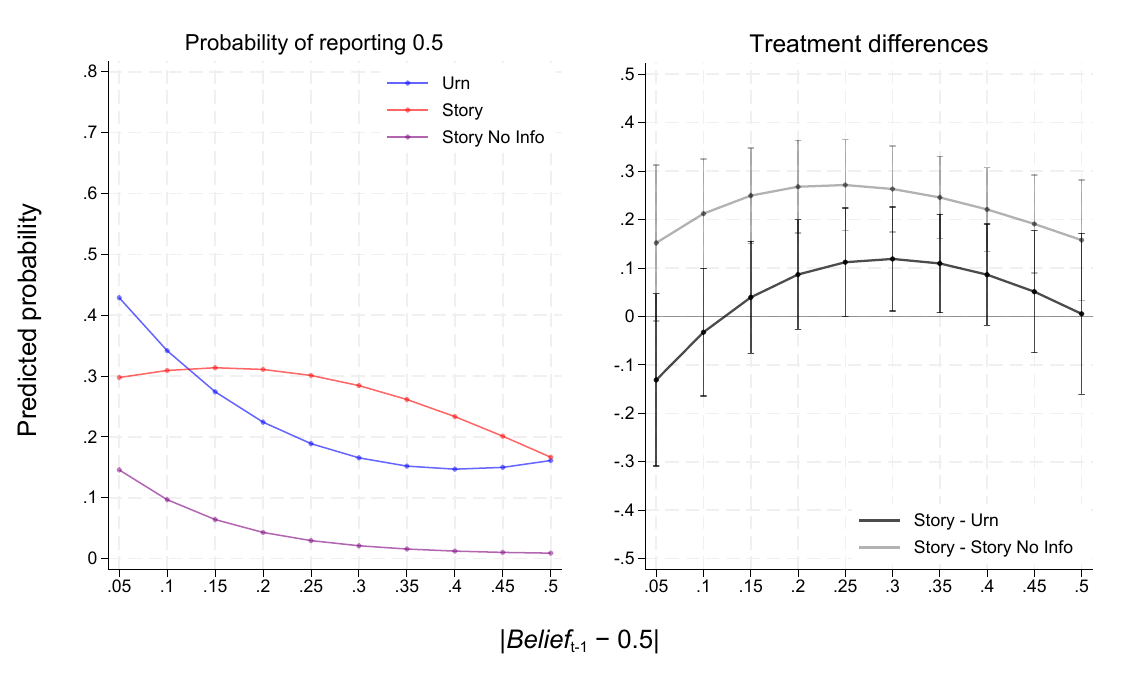}
    \caption{Predicted probability of exact belief reset (quadratic specification).\\{\footnotesize\textit{Note:} The left panel shows the predicted probability of reporting a belief of exactly 0.5 after observing a nondiagnostic signal for all treatment conditions. Predictions are obtained from a mixed-effects probit model that includes a quadratic polynomial in the absolute distance of the previous-round belief from 0.5, fully interacted with treatment. The right panel shows the predicted treatment differences for the relevant treatment comparisons. Vertical bars represent 95\% confidence intervals.}}\label{fig:prob_midpoint_quad}
\end{figure}

As an additional robustness check, we relax the linear functional-form assumption by including a quadratic in $d_{i,t-1}$, fully interacted with treatment. Figure~\ref{fig:prob_midpoint_quad} reports the resulting predicted probabilities of an exact belief reset and the corresponding treatment effects.

\begin{figure}[H]
    \centering
\includegraphics[width=1\linewidth]{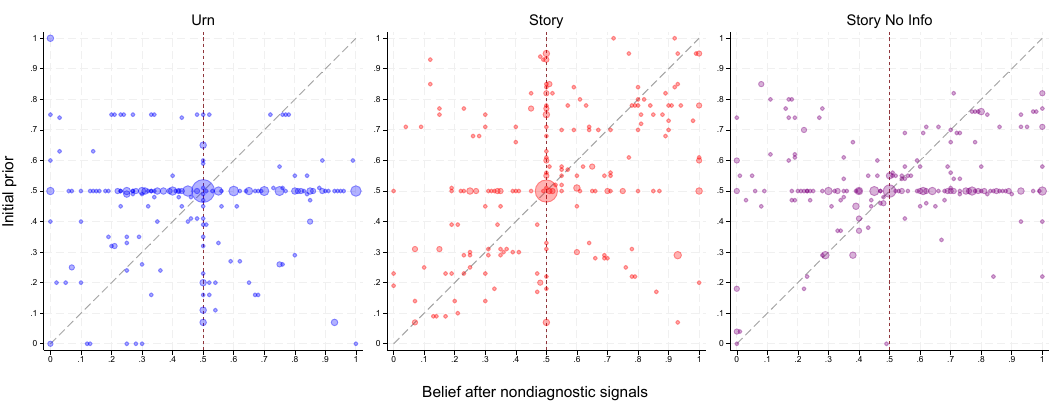}
    \caption{Initial prior and reports after nondiagnostic signals.\\{\footnotesize \textit{Note}: Each tile shows the reported belief following a nondiagnostic signal (x-axis) and the initial prior (y-axis) for each treatment. We exclude observations where the preceding belief is equal to 0.5. The size of the marker indicates the number of data points for that specific coordinate, where the smallest marker refers to one observation only.}} \label{fig:bubble_return_prior}
\end{figure}

\subsection{Pre-registered regression}\label{app:pre-reg}

The analysis presented in this section was included in the pre-registration and was intended to test the original hypothesis that nondiagnostic signals enter into the updating process to confirm prior beliefs. Because our data do not support such a hypothesis, we do not include the analysis in the paper and instead report it here for transparency. We follow \cite{CHARNESS20171} and \cite{KierenWeber2025}. Imagine the two states of the world are either Red (R state) or Green (G state).
Consider the $i_{th}$ person reporting the probability of the R state at each point in a sequence of 10 rounds. If she believes that both states are initially equally likely ex ante should report a 0.5 probability for the R state as an initial prior. Further, a Bayesian would calculate the probability of the R state after round t as (assuming she disregards non-informative signals)

\[
P_t^{\text{Bayes}} = P(\text{R} \mid z_t)^{\text{Bayes}} = \frac{\theta^{z_t}}{\theta^{z_t} + (1 - \gamma - \theta)^{z_t}}, \quad z_t = r_t - g_t
\]

where $r_t (g_t)$ denotes the number of Red (Green) signals that have been received as of round $t$. In our experiment, the proportion of Red's signals in the R state is $\theta = 0.45$, and the proportion of non-informative signals is $\gamma = 0.25$. Note that a Bayesian is indifferent regarding the order of the draws, since only the difference ($z_t$) is relevant. Thus, the natural log of the odds ratio for a Bayesian is given by

\[
\pi_t = \ln\left( \frac{P(\text{R} \mid z_t)^{\text{Bayes}}}{P(\text{G} \mid z_t)^{\text{Bayes}}} \right) 
= \ln\left( \frac{\theta}{1 - \gamma - \theta} \right) \times z_t = 0.4055 \times z_t \ \frac{<}{>} \ 0,
\]

and it is linear in info received; that is, the Bayesian log-odds ratio is updated by $\pm  0.4055 \times z_t$ after each new draw.\\
First-differencing both sides of the equation yields:

\[
\Delta \pi_t = \pi_t - \pi_{t-1} = \ln\left( \frac{\theta}{1 - \gamma - \theta} \right) \times \Delta z_t = 0.4055 \times \Delta z_t,
\]

where $\Delta \pi_t \in \{-0.4055, 0, 0.4055\}$ and $\Delta z_t \in \{-1, 0,  1\}$. In contrast to \cite{CHARNESS20171} and as in \cite{KierenWeber2025}, $z_t$ can take the value zero because of the nondiagnostic signals.\\
To isolate a ``confirming'' information condition, we construct the following dummy variables:

\[
C_t^R =
\begin{cases}
1 & \text{if } \pi_{t-1} > 0 \text{ and } s_t = r \\
0 & \text{otherwise}
\end{cases}
\]

\[
C_t^G =
\begin{cases}
1 & \text{if } \pi_{t-1} < 0 \text{ and } s_t = g \\
0 & \text{otherwise}
\end{cases}
\]

where $s_t$ is the signal received in the current round. These variables measure whether a Bayesian receiver would
view a received diagnostic signal as confirming a belief. For example, if a Bayesian believed that the R state of the world was more likely to be in play (i.e., $\pi_{t-1} > 0$) for a given draw $t$ and she receives a Red signal ($s_t = r$), then we say the belief of a
R state of the world is confirmed ($C_t^R = 1$). The same holds if that Bayesian believed that the G state of the world was more likely to be in play ($\pi_{t-1} < 0$) and she received a Green signal ($s_t = g$), in which case we would have that $C_t^G = 1$.\\
To account for cases where nondiagnostic signals are received ($s_t = n$), we construct the following dummy variables:

\[
C_t^{n_0} =
\begin{cases}
1 & \text{if } \pi_{t-1} = 0 \text{ and } s_t = n \\
0 & \text{otherwise}
\end{cases}
\]

\[
C_t^{n_R} =
\begin{cases}
1 & \text{if } \pi_{t-1} > 0 \text{ and } s_t = n \\
0 & \text{otherwise}
\end{cases}
\]

\[
C_t^{n_G} =
\begin{cases}
1 & \text{if } \pi_{t-1} < 0 \text{ and } s_t = n \\
0 & \text{otherwise}
\end{cases}
\]

$C_t^{n_R}$ and $C_t^{n_G}$ are variables that measure whether a Bayesian would view a nondiagnostic signal as confirming a belief, while $C_t^{n_0}$ accounts for events where the agent had equal prior on both states. Given the change in log-odds ($\Delta \pi_t$), and the dummies above, consider the following regression for a
Bayesian:

\[
\pi_{t} = \rho \pi_{t-1} + \beta \Delta z_{t} + \delta_1 C^R_{t} + \delta_2 C^G_{t} + \delta_3 C_t^{n_0} + \delta_4 C_t^{n_R} + \delta_5 C_t^{n_G} + \varepsilon_{t}
\]

Because a Bayesian would not be subject to either the conservatism/overreaction heuristic, nor would she place any additional weight on confirming diagnostic signals, or regard nondiagnostic ones, it must be the case that the coefficients satisfy

\[
\rho = 1, \quad \beta = \ln\left( \frac{\theta}{1 - \lambda - \theta} \right), \quad \delta_i = 0, \quad i = 1, 2, 3, 4, 5.
\]

\bigskip
\bigskip
\bigskip

\textbf{Actual Estimation}\\
The natural logarithm of an individual subject's odds ratio (the analog of $\pi_t$) within a round, based on her stated probability at each round $t$, that is $P_{it} = P_{it}(\text{R} \mid z_t)$, is:

\begin{equation}
\lambda_{it} = \ln(\Lambda_{it}) = \ln\left( \frac{P_{it}(\text{R} \mid z_t)}{1 - P_{it}(\text{R} \mid z_t)} \right)
\end{equation}

and may differ from $\pi_t$. To ensure that $\lambda_{it}$ is always defined, extreme beliefs of 0 and 1 are replaced with 0.01 and 0.99, respectively. 
As before, we construct the dummy variables
\[
C_{it}^R =
\begin{cases}
1 & \text{if } \lambda_{it-1} > 0 \text{ and } s_t = r \\
0 & \text{otherwise}
\end{cases}
\]

\[
C_{it}^G =
\begin{cases}
1 & \text{if } \lambda_{it-1} < 0 \text{ and } s_t = g \\
0 & \text{otherwise}
\end{cases}
\]

\[
C_{it}^{n_0} =
\begin{cases}
1 & \text{if } \lambda_{it-1} = 0 \text{ and } s_t = n \\
0 & \text{otherwise}
\end{cases}
\]

\[
C_{it}^{n_R} =
\begin{cases}
1 & \text{if } \lambda_{it-1} > 0 \text{ and } s_t = n \\
0 & \text{otherwise}
\end{cases}
\]

\[
C_{it}^{n_G} =
\begin{cases}
1 & \text{if } \lambda_{it-1} < 0 \text{ and } s_t = n \\
0 & \text{otherwise}
\end{cases}
\]

If subjects deviate from Bayesian behavior due to conservatism (over-reaction), one would expect to see log-odds for people that are consistently smaller (larger) than $\pm0.4055 \times z_t$. If people do weigh diagnostic evidence that confirms a previously held belief, or weigh nondiagnostic signals, then the dummies would predict log-odds. Thus, the regression for a subject, analogous to the one for a Bayesian, would be:

\[
\lambda_{it} = \rho \lambda_{it-1} + \beta \Delta z_{t} + \delta_1 C^R_{it} + \delta_2 C^G_{it} + \delta_3 C_{it}^{n_0} + \delta_4 C_{it}^{n_R} + \delta_5 C_{it}^{n_G} + \varepsilon_{it}
\]

clustering errors at the subject level, with a test of Bayesian behavior being that the estimate for $\rho$ equals 1, the estimate for $\beta$ equals $\ln\left(\frac{\theta}{1 - \gamma -\theta}\right)$, and that the coefficients on the dummies $C^R_{it}$, $C^G_{it}$, $C_{it}^{n_0}$, $C_{it}^{n_R}$ and $C_{it}^{n_G}$, namely, $\delta_1$, $\delta_2$, $\delta_3$, $\delta_4$, and $\delta_5$ be zero. Conservatism would manifest with an estimate of $\beta$ less than $\ln\left(\frac{\theta}{1 - \gamma - \theta}\right)$ whereas use of the over-reaction heuristic would imply an estimate of $\beta$ greater than $\ln\left(\frac{\theta}{1 - \gamma - \theta}\right)$. If subjects are affected by confirmation bias in that they place extra weight on a diagnostic signal that confirms beliefs, then it would be the case that the estimates of $\delta_1$ and $\delta_2$ would be nonzero. 
Nonzero coefficients of $\delta_3$, $\delta_4$, and $\delta_5$ capture sign-specific deviations from the belief persistence implied by $\rho$. Because the total predicted belief change depends jointly on $\rho$ and these coefficients, the latter should not be interpreted in isolation. In summary, a test of Bayesian behavior would be:

\[
H_0 : \hat{\rho} = 1 \quad \text{and} \quad \hat{\beta} = 0.4055 \quad \text{and} \quad \hat{\delta}_i = 0, \quad i = 1, 2, 3 ,4 ,5.
\]

The estimation results are presented in Table \ref{tab:pre-registered}, with the regression estimated separately for each treatment.

\begin{table}[H]\centering
\def\sym#1{\ifmmode^{#1}\else\(^{#1}\)\fi}
\caption{Pre-registered regressions.\label{tab:pre-registered}}
\begin{tabular}{l*{3}{cc}}
\hline\hline
          &\multicolumn{1}{c}{\Narrative}&\multicolumn{1}{c}{\NarrativeNI}&\multicolumn{1}{c}{\Control}\\
\hline
$\rho$     &0.580\sym{***} (0.06)         &0.742\sym{***} (0.05)         &0.620\sym{***} (0.06)         \\
$\beta$   &0.898\sym{***} (0.08)         &0.654\sym{***} (0.06)         &0.665\sym{***} (0.07)         \\
$\delta_1$ &0.294\sym{*} (0.15)         &0.242\sym{*} (0.10)         &0.475\sym{***} (0.13)         \\
$\delta_2$ &-0.349\sym{*} (0.17)         &-0.297\sym{*} (0.15)         &-0.442\sym{**} (0.16)         \\
$\delta_3$ &-0.057 (0.07)         &-0.121 (0.08)         &0.042 (0.03)         \\
$\delta_4$ &-0.081 (0.12)         &0.181\sym{*} (0.09)         &0.037 (0.09)         \\
$\delta_5$ &0.144 (0.14)         &-0.291\sym{**} (0.09)         &0.183 (0.13)         \\
\hline
$H_0: \rho = 1$ & $<$0.001 & $<$0.001 & $<$0.001 \\
$H_0: \beta = 0.4055$ & $<$0.001 & $<$0.001 & $<$0.001 \\
$H_0:$ Bayesian & $<$0.001 & $<$0.001 & $<$0.001 \\
\hline
Observations& 1,152         & 1,008         & 1,251         \\
\hline\hline
\end{tabular}
\caption*{\footnotesize\textit{Note:} Standard errors clustered at the subject level are in parentheses.\footnotesize \sym{\dagger} \(p<0.10\), \sym{*} \(p<0.05\), \sym{**} \(p<0.01\), \sym{***} \(p<0.001\)}
\end{table}

Table~\ref{tab:pre-registered} rejects the joint Bayesian restrictions in all three treatments. The estimate of $\rho$ is significantly below one throughout, implying contraction of reported log-odds toward 0.5, while the estimate of $\beta$ is significantly above its Bayesian benchmark, indicating overreaction to diagnostic signals. The latter result is qualitatively consistent with the reaction-intensity estimates reported in Appendix~\ref{app:types}, which exceed one in all treatments. The positive estimates of $\delta_1$ and negative estimates of $\delta_2$ indicate that diagnostic signals receive additional weight when they confirm the direction of the previous belief. By contrast, the preregistered regression provides no evidence that explicitly presented nondiagnostic clues reinforce prior beliefs in \Narrative. In this treatment, the estimates of $\delta_4$ and $\delta_5$ are insignificant and have signs consistent with reset to 0.5. Both coefficients are also insignificant in \Control. In \NarrativeNI, the estimated $\delta_4$ is positive and $\delta_5$ negative, and both are significant. Because $\hat{\rho}<1$, however, these coefficients partly counteract the contraction implied by the estimate of $\rho$ and should not be interpreted in isolation as movement away from 0.5. Overall, the estimates reject Bayesian updating and do not support the original hypothesis that nondiagnostic narrative clues confirm existing beliefs.

\newpage

\section{Behavioral Types}\label{app:types}

Let $b_{t-1} \in [0,1]$ denote the subject's prior belief held at the start of round $t$. Let $s_t \in \{-1,0,+1\}$ denote the realized signal, where $s_t=+1$ favors the true state, $s_t=-1$ disfavors it, and $s_t=0$ is nondiagnostic. For each of the four candidate rules $k$, we define a rule-specific predicted belief $m_t^k$. Reported beliefs are modeled as noisy realizations around this prediction, with support restricted to the feasible belief interval:
\[
b_t \mid k \sim \mathcal{N}_{[0,1]}\left(m_t^k,\sigma_k^2\right),
\]
where \(\mathcal{N}_{[0,1]}(m_t^k,\sigma_k^2)\) denotes a normal distribution with location parameter \(m_t^k\) and scale parameter \(\sigma_k\), truncated to the interval \([0,1]\). In estimation, we restrict the scale parameter to \(\sigma_k\in[0.0001,0.5]\). Each type is therefore characterized by a deterministic updating rule and a rule-specific noise parameter.\\

\subsection*{The four updating rules}

\paragraph{Bayes.}
This rule preserves the direction of Bayesian updating while allowing the strength of the reaction to diagnostic signals to differ from the Bayesian benchmark. In log-odds form,
\[
\log\!\left(\frac{m_t^{RI}}{1-m_t^{RI}}\right)
=
\log\!\left(\frac{b_{t-1}}{1-b_{t-1}}\right)
+
\lambda\,\ell(s_t),
\]
where
\[
\ell(s_t)=
\begin{cases}
\log\!\left(\frac32\right), & \text{if } s_t=+1, \\[4pt]
0, & \text{if } s_t=0, \\[4pt]
\log\!\left(\frac23\right), & \text{if } s_t=-1.
\end{cases}
\]
Equivalently,
\[
m_t^{RI}=
\begin{cases}
\dfrac{b_{t-1}\left(\frac32\right)^{\lambda}}
{b_{t-1}\left(\frac32\right)^{\lambda}+(1-b_{t-1})},
& \text{if } s_t=+1, \\[12pt]
b_{t-1}, & \text{if } s_t=0, \\[10pt]
\dfrac{b_{t-1}\left(\frac23\right)^{\lambda}}
{b_{t-1}\left(\frac23\right)^{\lambda}+(1-b_{t-1})},
& \text{if } s_t=-1.
\end{cases}
\]
The parameter $\lambda$ measures the intensity of reaction to diagnostic signals. When $\lambda=1$, the rule coincides with Bayesian updating under diagnostic signals; when $0<\lambda<1$, the subject underreacts; and when $\lambda>1$, the subject overreacts.\footnote{This specification is a parsimonious reduced-form distortion of Bayesian updating in log-odds form. It is closely related to Grether's non-Bayesian odds-ratio specification and to more recent empirical work that models departures from Bayes directly in log-odds form. Our setup can be read as the special case in which the coefficient on prior log-odds is fixed at one and the coefficient on the signal log-likelihood ratio is $\lambda$. For related approaches and for a characterization of Bayesian updating itself \citep[see][]{grether1980bayes,coutts2019good,alosferrermihm2023bayesian}.}

\paragraph{Bayes-Reset.}
The second model has the same smooth response to diagnostic signals, but resets beliefs to 0.5 after a nondiagnostic one:
\[
m_t^{RIR}=
\begin{cases}
\dfrac{b_{t-1}\left(\frac32\right)^{\lambda}}
{b_{t-1}\left(\frac32\right)^{\lambda}+(1-b_{t-1})},
& \text{if } s_t=+1, \\[12pt]
\dfrac12, & \text{if } s_t=0, \\[10pt]
\dfrac{b_{t-1}\left(\frac23\right)^{\lambda}}
{b_{t-1}\left(\frac23\right)^{\lambda}+(1-b_{t-1})},
& \text{if } s_t=-1.
\end{cases}
\]
Hence, the first two models differ only in their treatment of nondiagnostic information. In the first case, a nondiagnostic signal leaves beliefs unchanged, while in the second case, it induces a belief reset to 0.5. This reset component is motivated by our results and evidence that individuals often respond in a non-Bayesian way to weak or nondiagnostic signals \citep[e.g.,][]{aydogan2017signal,KierenWeber2025}.

\paragraph{Coarse.}
The third model reacts in the most extreme possible way to diagnostic signals, but leaves beliefs unchanged after a nondiagnostic one:
\[
m_t^{EI}=
\begin{cases}
1, & \text{if } s_t=+1, \\[4pt]
b_{t-1}, & \text{if } s_t=0, \\[4pt]
0, & \text{if } s_t=-1.
\end{cases}
\]

\paragraph{Coarse-Reset.}
The fourth model also reacts in the most extreme possible way to the sign of diagnostic signals and triggers a belief reset to 0.5 after a nondiagnostic one:
\[
m_t^{ER}=
\begin{cases}
1, & \text{if } s_t=+1, \\[4pt]
\dfrac12, & \text{if } s_t=0, \\[4pt]
0, & \text{if } s_t=-1.
\end{cases}
\]
Thus diagnostic signals generate corner beliefs, while nondiagnostics generate a return to \(\frac12\).

Hence the difference between the two extreme models mirrors the difference between the two Bayesian models: belief reset to 0.5 versus no updating following nondiagnostic signals.

\subsection*{BIC-based classification}

For each subject, each candidate model is fit separately using all available updating rounds for that subject. Conditional on model $k$, observed beliefs are assumed to be independent draws from the truncated-normal distribution defined above, centered at the rule-implied values $m_t^k$. If a subject contributes observations indexed by $t=1,\dots,T_i$, the log-likelihood of model $k$ is
\[
\log L_i^{(k)}
=
\sum_{t=1}^{T_i}
\log\!\left[
\frac{
\phi\!\left(\frac{b_t-m_t^k}{\sigma_k}\right)
}{
\sigma_k
\left[
\Phi\!\left(\frac{1-m_t^k}{\sigma_k}\right)
-
\Phi\!\left(\frac{-m_t^k}{\sigma_k}\right)
\right]
}
\right],
\]
where $\phi(\cdot)$ and $\Phi(\cdot)$ denote the standard normal density and cumulative distribution function, respectively.

Equivalently,
\[
\log L_i^{(k)}
=
\sum_{t=1}^{T_i}
\left[
\log \phi\!\left(\frac{b_t-m_t^k}{\sigma_k}\right)
-
\log \sigma_k
-
\log\!\left(
\Phi\!\left(\frac{1-m_t^k}{\sigma_k}\right)
-
\Phi\!\left(\frac{-m_t^k}{\sigma_k}\right)
\right)
\right].
\]

The two Bayesian models each have two free parameters, a noise parameter $\sigma_k$ and a reaction-intensity parameter $\lambda$. The two Coarse models only estimate the noise parameter $\sigma_k$. To avoid mechanically favoring the more flexible specifications, we classify subjects using the Bayesian Information Criterion {\citep{schwarz1978estimating}}:
\[
\mathrm{BIC}_i^{(k)} = -2\log L_i^{(k)} + p_k \log T_i,
\]
where $p_k$ is the number of free parameters in model $k$. Thus $p_k=2$ for the two Bayesian models and $p_k=1$ for the two Coarse models. Each subject is assigned to the model with the lowest BIC,
\[
\hat{k}_i = \arg\min_{k \in \{B,BR,C,CR\}} \mathrm{BIC}_i^{(k)}.
\]
Hence, the classification compares smooth and extreme reactions to diagnostic signals while allowing for either belief reset or no update after nondiagnostic signals, but penalizes the two Bayesian models for their additional parameter.

\subsection*{Additional Results}

Across treatments, the fitted reaction-intensity parameters ($\lambda$) exceed one, indicating overreaction to diagnostic signals. This is consistent with the preregistered regressions presented in Section~\ref{app:pre-reg}, where the estimated $\beta$ is significantly larger than its Bayesian benchmark in every condition.

\begin{table}[htbp]
\centering\small
\caption{Estimated reaction-intensity parameter $(\lambda)$.}
\label{tab:lambda_truncnorm}
\begin{tabular}{lccc}
\toprule
 & \Control & \Narrative & \NarrativeNI \\
\midrule
Bayes
  & 2.062 [1.726, 2.477] & 3.282 [2.344, 4.442] & 1.678 [1.365, 2.097] \\
Bayes-Reset
  & 2.053 [1.421, 2.827] & 2.080 [1.613, 2.604] & 2.072 [1.307, 2.926] \\
\midrule
\addlinespace[+15pt]

 & \Narrative $-$ \Control & \NarrativeNI $-$ \Control & \Narrative $-$ \NarrativeNI \\
 \midrule
Bayes
  & 1.208 [0.199, 2.425] & $-0.383$ [$-0.898$, 0.150] & 1.594 [0.581, 2.804] \\
  & (0.019) & (0.153) & (0.001) \\
Bayes-Reset
  & 0.022 [$-0.884$, 0.846] & 0.011 [$-1.048$, 1.084] & 0.007 [$-0.942$, 0.939] \\
  & (0.961) & (0.983) & (0.989) \\
\bottomrule
\end{tabular}
\caption*{\footnotesize\textit{Note:} The top panel reports bootstrap medians of the average estimated $\lambda$ among subjects classified as either Bayes or Bayes-Reset. The bottom panel reports treatment differences. Percentile 95\% confidence intervals are reported in brackets. Two-sided bootstrap $p$-values for differences between treatments are reported in parentheses.}
\end{table}

\begin{table}[H]
\centering\small
\caption{Estimated $\sigma$ and upper-bound diagnostics.}
\label{tab:sigma_truncnorm}
\begin{tabular}{lccc}
\toprule
\multicolumn{4}{l}{\emph{Panel A. Median $\widehat\sigma$ among subjects assigned to each type}}\\
\midrule
 & \Control & \Narrative & \NarrativeNI \\
\cline{2-4}
Bayes
  & 0.085 [0.074, 0.097] & 0.122 [0.101, 0.145] & 0.096 [0.085, 0.107] \\
Bayes-Reset
  & 0.094 [0.073, 0.121] & 0.125 [0.108, 0.149] & 0.135 [0.089, 0.186] \\
Coarse
  & 0.246 [0.129, 0.348] & 0.329 [0.243, 0.417] & 0.260 [0.226, 0.296] \\
Coarse-Reset
  & 0.383 [0.319, 0.440] & 0.229 [0.187, 0.272] & 0.413 [0.338, 0.500] \\
\addlinespace
\midrule
\multicolumn{4}{l}{\emph{Panel B. Fraction of selected fits with $\widehat\sigma$ near 0.5}} \\
\midrule
 & \Control & \Narrative & \NarrativeNI \\
 \cline{2-4}
Bayes
  & 0.000 [0.000, 0.000] & 0.000 [0.000, 0.000] & 0.000 [0.000, 0.000] \\
Bayes-Reset
  & 0.000 [0.000, 0.000] & 0.000 [0.000, 0.000] & 0.000 [0.000, 0.000] \\
Coarse
  & 0.000 [0.000, 0.000] & 0.250 [0.000, 0.600] & 0.000 [0.000, 0.000] \\
Coarse-Reset
  & 0.263 [0.063, 0.500] & 0.000 [0.000, 0.000] & 0.333 [0.000, 1.000] \\
\bottomrule
\end{tabular}
\caption*{\footnotesize\textit{Note:} Panel A reports the bootstrap medians of the estimated $\sigma$ among subjects assigned to each type. Panel B reports the fraction of those selected fits with $\widehat\sigma\geq0.4995001$. The admissible range is $[0.0001,0.5]$. Percentile 95\% confidence intervals are reported in brackets.}
\end{table}

\end{document}